\documentclass[twocolumn,prd,showpacs]{revtex4}
\usepackage{graphicx}% Include figure files

\begin{document}

\title{Torsion Cosmology and the Accelerating Universe}

\author{Kun-Feng Shie$^1$}
\author{James M. Nester$^{1,2,3}$}
\email{nester@phy.ncu.edu.tw}
\author{Hwei-Jang Yo$^4$}
\email{hjyo@phys.ncku.edu.tw}
\affiliation{$^1$Department of Physics,
 National Central University, Chungli 320, Taiwan}
\affiliation{$^2$Graduate Institute of Astronomy,
 National Central University, Chungli 320, Taiwan}
 \affiliation{$^3$Center for Mathematics and Theoretical Physics,
National Central University, Chungli 320, Taiwan}
\affiliation{$^4$Department of Physics, National Cheng-Kung
University,
 Tainan 701, Taiwan}

\date{\today}

%%%%%%%%%%%%%%%%%%%%%%%%%%%%%%%%%%%%%%%%%%%%%%%%%%%%%
\begin{abstract}
%%%%%%%%%%%%%%%%%%%%%%%%%%%%%%%%%%%%%%%%%%%%%%%%%%%%%
Investigations of the dynamic modes of the Poincar{\'e} gauge theory
of gravity found only two good propagating torsion modes;
they are effectively a scalar and a pseudoscalar.
Cosmology affords a natural situation where one might see observational
effects of these modes.
Here we consider only the ``scalar torsion'' mode.
This mode has certain distinctive and interesting qualities.
In particular this type of torsion does not interact directly with any known matter
and it allows a critical non-zero value for the affine scalar curvature.
Via numerical evolution of the coupled nonlinear equations we show that
this mode can contribute an oscillating aspect to the expansion rate
of the Universe.  From the examination of specific cases of the
parameters and initial conditions we show that for suitable ranges
of the parameters the dynamic ``scalar torsion'' model can display
features similar to those of the presently observed accelerating
universe.
\end{abstract}

\maketitle

%%%%%%%%%%%%%%%%%%%%%%%%%%%%%%%%%%%%%%%%%%%%%%%%%%%%%
\section{Introduction}
%%%%%%%%%%%%%%%%%%%%%%%%%%%%%%%%%%%%%%%%%%%%%%%%%%%%%
One of the outstanding successes of theoretical physics in the
latter part of the last century which led to a much deepened
understanding was the recognition that all the known fundamental
physical interactions, the strong, weak, and electromagnetic---{\it
not excepting gravity}---can be well described in terms of a single
unifying principle: that of local gauge theory. Although there are
other possible gauge approaches, for gravity it seems highly
appropriate to regard it a gauge theory for the local symmetry group
of Minkowski space time: the Poincar\'e group \cite{HHKN,Nes84}.
Such a consideration led to the development of the  Poincar{\'e}
Gauge Theory of gravity (PGT)
\cite{Hehl80,HS80,MieE87,HHMN95,GFHF96,Blag02}. The PGT has {\em a
priori} independent local rotation and translation potentials, which
correspond to the metric-compatible connection and orthonormal
co-frame; their associated field strengths are the {\em curvature}
and {\em torsion}. The spacetime then has generically a
Riemann-Cartan geometry. Because of its gauge structure and
geometric properties the PGT has been regarded as an attractive
alternative to general relativity. The general theory includes as
exceptional cases Einstein's general relativity (GR) with {\em
vanishing} torsion, the Einstein-Cartan theory with {\em
non-dynamic} torsion algebraically coupled to the intrinsic spin of
the source, as well as the teleparallel theories wherein curvature
vanishes and torsion represents the gravitational force (a sort of
opposite to Riemannian geometry). Aside from these exceptions the
generic PGT has, in addition to the metric familiar from Einstein's
GR, a connection with some independent dynamics.  This additional
dynamics is reflected in the torsion tensor.

There is a natural physical source for the torsion of spacetime:
namely spin $1/2$ fermions. The effect is generally assumed to be
small at ordinary densities, but could have a major influence at
high densities (e.g., beyond $10^{48}$ gm/cm${}^3$), and thus it was
expected to have important physical effects in the early universe
\cite{HHKN}. Torsion cosmology investigations were initiated
by Kopc{\'n}yski \cite{Kop}. Some early investigations attracted
attention especially because they noted that torsion might prevent
the (at that time newly recognized) singularities. However, this
hope quickly faded.  Indeed it soon was argued that {\it non-linear}
torsion effects were more likely to produce stronger singularities
\cite{IN77}.

The various PGT dynamic modes beyond those of the metric were first
investigated via the linearized theory (for outstanding examples of
such investigations see \cite{HS80,SN80}). To this order the
connection dynamics (which can be represented by the torsion tensor)
decomposes into six modes with certain spins and parity:
$2^\pm,1^\pm,0^\pm$. Many possible combinations of well behaved
(carrying positive energy at speed $\le c$, criterion often referred
to as ``no ghost, no tachyon'') propagating modes
 in the linear PGT theory
were identified. They were classified into about a dozen separate
cases, almost any combination of up to 3 dynamic modes is allowed.
Some nice investigations of the PGT theory were also made using the
Hamiltonian analysis \cite{Blag02,BMNI83,NicI84}, with findings
consistent with the conclusions of the linearized investigation.
Later, however, some potential problems were identified
\cite{HRLW91}. This prompted deeper investigations, which
 noted that effects due to non-linearities in the constraints could
be expected to render most of the aforementioned dynamic cases
physically unacceptable \cite{CNY98}. A fundamental investigation
identified two special cases, the so-called ``scalar torsion''
modes, which could be proved to be problem free, having a well posed
initial value problem \cite{HNZ96}. Subsequently Hamiltonian
investigations
 \cite{YN99,YN02} supported the conclusion that these two dynamic
``scalar torsion'' modes may well be the only physically acceptable
dynamic PGT torsion modes.

In one mode (referred to as the ``pseudoscalar'' because of its
$0^-$ spin content) only the axial vector torsion is dynamic. (As a
consequence of the dynamic field equations it turns out to be
 dual to the gradient of a scalar field; however it is not possible to
 treat
this scalar field as the primary dynamical object without changing
the nature of the theory \cite{0minus}). Axial torsion is naturally
driven by the intrinsic spin of fundamental fermions; in turn it
naturally interacts with such sources. Thus for this mode one has
some observational constraints \cite{CSFG94}. Note that except in
the early universe one does not expect large spin densities.
Consequently it is generally thought that axial torsion must be
small and have small effects at the present time. This is one reason
why we do not focus on this mode here.

The other good mode, $0^+$, the so-called ``scalar torsion'' mode,
has a certain type of dynamic vector torsion. (As a consequence of
the dynamic equations given below in \S II.B, it too turns out to be
the gradient of a scalar field; this scalar field cannot, however,
be regarded as a fundamental potential---for essentially the same
reasons as those mentioned in connection with the ``pseudoscalar''
mode). There is no known fundamental source which directly excites
this mode. Conversely this type of torsion does not interact in any
direct obvious fashion with any familiar type of matter
\cite{ShaI02}. Hence we do not have much in the way of constraints
as to its magnitude.  We could imagine it as having significant
magnitude and yet not being dramatically noticed---except indirectly
through the non-linear equations.  This mode in particular has also
attracted our interest because of a conspicuous consequence of the
non-linear equations: in this case there is a critical non-zero
value for the affine scalar curvature.

Our theoretical PGT analysis thus led us to consider just two
dynamic torsion modes. An obvious place where we might see some
physical evidence for these modes is in cosmological models.  The
homogeneous and isotropic assumptions of cosmology greatly restrict
the possible types of non-vanishing fields.  Curiously, for torsion
there are only two possibilities: $0^+$, i.e., vector torsion which,
moreover, has only a time component (and is thus effectively the
gradient of a time-dependent function), and axial torsion, $0^-$,
which is effectively the dual of a vector with only a time component
(and thus can be specified as the gradient of a time-dependent
function). Hence the homogeneous and isotropic cosmologies are {\em
naturally} very suitable for the exploration of the physics of the
dynamic PGT ``scalar modes''.

Thus cosmological models  offer a situation where dynamic torsion
may lead to observable effects.  Here we will not focus on the early
universe, where one could surely expect large effects (although
their signature would have to be disentangled from other large
effects), and instead inquire whether one can see traces of torsion
effects today.  In particular we will here consider accounting for
the outstanding present day mystery: the accelerated universe, in
terms of an alternate gravity theory with an additional natural
dynamic geometric quantity: torsion \cite{YN06}.

The observed accelerating expansion of the Universe suggested the
existence of a kind of dark energy with a negative pressure. The
idea of a dark energy is one of the greatest challenges for our
current understanding of fundamental physics
\cite{PPRB03,PadT03,CoST06}. Among a number of possibilities to
describe this dark energy component, the simplest may well be by
means of a cosmological constant $\Lambda$. However, there are some
reasons for dissatisfaction with this model. In particular the
so-called {\em {the cosmological constant problem}} notes that the
theoretically estimated value of the vacuum energy density is about
$10^{120}$ times larger than the inferred cosmological constant.
Moreover the {\em coincidence} or {\em fine-tuning} problem notes
that it is highly unlikely that we should be living in the
relatively short era when the rapidly changing ratio of the material
energy and the cosmological constant is nearly unity.

In the light of these problems there have been many interesting
dynamical dark energy proposals. A popular idea is some unusual type
of minimally coupled scalar field $\Phi$ (quintessence field) which
has not yet reached its ground state and whose current dynamics is
basically determined by its potential energy $V(\Phi)$. This idea
has received much attention over the past few years and a
considerable effort has been made in understanding the role of
quintessence fields on the dynamics of the Universe (see, e.g.,
\cite{CaDS98,CarS98,BKBR05}). However, without a specific motivation
from fundamental physics for the light scalar fields, these
quintessence models can be constructed relatively arbitrarily. There
is a lot of room for speculation.

Here we consider another possibility for explaining the accelerating
universe: dynamic scalar torsion. We explore the possibility that
the dynamic PGT connection, reflected in dynamic PGT torsion,
provides the accelerating force in the universe. As noted above,
there are certain ``scalar torsion'' modes which could have
dynamical behavior. They could naturally provide the accelerating
force in the universe. Here we will show that the effect of torsion
can not only make the expansion rate oscillate, but also can force
the universe to naturally have an accelerating expansion in some
periods and a decelerating expansion at other times. Scalar torsion
cosmology can avoid some of the problems which occur in other
models.

A comprehensive survey of the PGT  cosmological models was presented
some time ago by Goenner and M{\"u}ller-Hoissen \cite{GMH}. Although
that work only solved in detail a few particular cases, it developed
the equations for all the PGT cases---including those for the
particular model we consider here. However that work was done prior
to the discovery of the accelerating universe, and torsion was thus
imagined as playing a big role only at high densities in the early
universe. More recently investigators have begun to consider torsion
as a possible cause of the accelerating universe (see,
e.g.,~\cite{Boe03,CaCT03}) but the subject has not yet been explored
in detail \cite{MGK07}.

We have taken a first step in the exploration of the possible
evolution of the Universe with the scalar torsion mode of the PGT.
The main motivation is two-fold: (1) to have a better understanding
of the PGT, in particular the possible physics of the dynamic
``scalar torsion'' modes; (2) to consider the prospects of
accounting for the outstanding present day mystery---the
accelerating universe---in terms of an alternate gravity theory,
more particularly in terms of the PGT dynamic torsion. With the
usual assumptions of isotropy and homogeneity in cosmology, we find
that, under the model, the Universe will oscillate with generic
choices of the parameters. The torsion field in the model plays the
role of the imperceptible ``dark energy''. With a certain range of
parameter choices, it can account for the  current status of the
Universe, i.e., an accelerating expanding universe with a value of
the Hubble constant which is approximately the present one. These
promising results should encourage further investigations of this
model, with a detailed comparison of its predictions with the
observational data.

The remainder of this work is organized as follows: We summarize the
formulation of the PGT and the ``scalar torsion'' mode in Sec.~II,
and then translate the equations into a certain effective Riemannian
form in Section~III; the specialization of these relations to the
form describing a cosmological model are presented in Sec.~IV. Then
 a preliminary analytical analysis aimed at revealing the behavior
of the solutions is presented in Sec.~V.
 In Section VI we present the results of our numerical
demonstrations for various choices of the parameters and the initial
data. The implications of our findings are discussed in Section VII
and Sec.~VIII is a conclusion.

%%%%%%%%%%%%%%%%%%%%%%%%%%%%%%%%%%%%%%%%%%%%%%%%%%%%%
\section{The Field Equations}
%%%%%%%%%%%%%%%%%%%%%%%%%%%%%%%%%%%%%%%%%%%%%%%%%%%%%
\label{secpgt}
%%%%%%%%%%%%%%%%%%%%%%%%%%%%%%%%%%%%%%%%%%%%%%%%%%%%%
\subsection{Poincar\'e gauge theory of gravitation}
%%%%%%%%%%%%%%%%%%%%%%%%%%%%%%%%%%%%%%%%%%%%%%%%%%%%%

Our considerations in this work are entirely classical. The form of
the gravity theory we wish to consider here, the PGT, was worked out
some time ago on the basis of the fundamental principles of gauge
theory and geometry
\cite{HHKN,Nes84,Hehl80,HS80,MieE87,HHMN95,GFHF96,Blag02}.
 In the PGT there are two sets of local gauge potentials, the orthonormal
 frame field (tetrad) $e_i{}^\mu$ and the metric-compatible connection
 $\Gamma_{i\mu}{}^\nu$, which are associated with the translation and the
Lorentz
 subgroups of the Poincar\'e gauge group, respectively.
 The
 field strengths associated with the frame and connection are the torsion
 \begin{equation}
  T_{ij}{}^\mu =2(\partial_{[i}e_{j]}{}^\mu
  +\Gamma_{[i|\nu }{}^\mu e_{|j]}{}^\nu),\label{torsiondef}
 \end{equation}
 and the curvature
 \begin{equation}\label{defR}
  R_{ij\mu}{}^\nu =2(\partial_{[i}\Gamma_{j]\mu}{}^\nu
  +\Gamma_{[i|\sigma}{}^\nu \Gamma_{|j]\mu}{}^\sigma),
 \end{equation}
 which satisfy the  Bianchi identities
 \begin{eqnarray}
  &&\nabla_{[i}T_{jk]}{}^\mu \equiv R_{[ijk]}{}^\mu,\\
  &&\nabla_{[i}R_{jk]}{}^{\mu \nu}\equiv 0.
 \end{eqnarray}
From the frame one constructs some auxiliary quantities: the
reciprocal frame $e^i{}_\mu$, which satisfies $e^i{}_\mu e_i{}^\nu
=\delta_\mu {}^\nu$ and $e^i{}_\mu e_j{}^\mu =\delta_j{}^i$, and the
metric $g_{ij}=e_i{}^\mu e_j{}^\nu \eta _{\mu \nu}$.
Here our conventions are as follows:
the Greek indices are the local Lorentz indices;
whereas the Latin indices are the coordinate indices.
We use the metric signature ($-$,$+$,$+$,$+$).

 Following the standard paradigm, the conventional form of the PGT action, which is invariant under
 local Poincar\'e gauge transformations, is taken to have the form
 \begin{equation}
  A=\int d^4x e(L_{\rm G}+L_{\rm M}).
 \end{equation}
 Here $e={\rm det}(e_i{}^\mu )$, $eL_{\rm G}(e_i{}^\mu,\partial_j e_i{}^\mu,\Gamma_{i\mu}{}^\nu,\partial_j \Gamma_{i\mu}{}^\nu)=
 eL_{\rm G}(e_i{}^\mu, T_{ij}{}^\mu, R_{ij}{}^{\mu\nu})$ is the geometric gravity Lagrangian density and
 $eL_{\rm M}(e,\Gamma,\psi,\partial\psi)=eL_{\rm M}(e_i{}^\mu, \psi, D_i\psi)$ is the minimally coupled source Lagrangian density,
 where $\psi$ represents all the matter and other interaction fields.
% (which
%determines the field equations for the source as well as the
%gravitational source currents, the material energy-momentum and spin
%density), $L_g$ denotes the gravitational Lagrangian density, and .
%In this paper we are concerned with the gravitational propagating
%modes,  we do not need the explicit equations of motion for the
%sources.
We will not explicitly need the field equations for the
non-geometric fields.  Varying with respect to the geometric-gauge
potentials gives the gravitational field equations.  As explained in
 detail in the aforementioned references, they take the form
 \begin{eqnarray}
  \nabla_jH_\mu{}^{ij}-E_\mu{}^i&=&{\cal T}_\mu{}^i,\label{first}\\
  \nabla_jH_{\mu \nu}{}^{ij}-E_{\mu
  \nu}{}^i&=&S_{\mu\nu}{}^i,\label{second}
 \end{eqnarray}
 with the field momenta
 \begin{eqnarray}
  H_\mu {}^{ij}&:=&{\partial e L_{\rm G}\over \partial\partial_j e_i{}^\mu}
  =2{\partial e L_{\rm G}\over \partial T_{ji}{}^\mu},\\
  H_{\mu \nu}{}^{ij}&:=&{\partial e L_{\rm G}\over
   \partial\partial_j\Gamma_i{}^{\mu \nu}}
  =2{\partial e L_g\over \partial R_{ji}{}^{\mu \nu}},
 \end{eqnarray}
 and
 \begin{eqnarray}
  E_\mu {}^i&:=&e^i{}_\mu e L_{\rm G}-T_{\mu j}{}^\nu H_\nu {}^{ji}
  -R_{\mu j}{}^{\nu\sigma}H_{\nu\sigma}{}^{ji},\\
  E_{\mu \nu}{}^i&:=&H_{[\nu \mu]}{}^i.
 \end{eqnarray}
The source terms here
\begin{eqnarray}
 {\cal T}_\mu{}^i&:=&\frac{\partial eL_{\rm M}}{\partial e_i{}^\mu},\\
S_{\mu\nu}{}^i&:=&
    \frac{\partial eL_{\rm M}}{\partial \Gamma_i{}^{\mu\nu}},
\end{eqnarray}
are, respectively, the Noether energy-momentum and spin density
currents, which (as a consequence of the assumed minimal coupling)
automatically satisfy suitable energy-momentum and angular momentum
conservation laws.

The Lagrangian is chosen (as usual in gauge theories) to be at most
of quadratic order in the field strengths,
 then the field momenta are
linear in the field strengths:
 \begin{eqnarray}
  H_\mu {}^{ij}&=&{e\over l^2}\sum^3_{k=1} a_k
  {\buildrel {(k)} \over T}{}^{ji}{}_\mu,\\
  H_{\mu \nu}{}^{ij}&=&-{a_0 e\over l^2}e^i{}_{[\mu}e^j{}_{\nu ]}
  +{e \over \kappa}\sum^6_{k=1}b_k
  {\buildrel {(k)} \over R}{}^{ji}{}_{\mu \nu};
 \end{eqnarray}
here the three $\displaystyle {{\buildrel {(k)} \over
T}{}^{ji}{}_\mu}$ and the six $\displaystyle { {\buildrel {(k)}
\over R}{}^{ji}{}_{\mu \nu}}$ are the algebraically irreducible
parts of the torsion and the curvature, respectively. The torsion in
particular splits into the algebraically irreducible torsion vector,
axial vector and tensor:
\begin{eqnarray}
 T_i&=&T_{ij}{}^j\,,\nonumber\\
 P_i&=&\frac{1}{2}\epsilon_{ijkm}T^{jkm}\,,\\
 Q_{ijk}&=&T_{i(jk)}-\frac{1}{3}T_ig_{jk}+\frac{1}{3}g_{i(j}T_{k)}\,,\nonumber
\end{eqnarray}
which recompose to give
\begin{equation}
 T_{ijk}=\frac{4}{3}Q_{[ij]k}+\frac{2}{3}T_{[i}g_{j]k}
        +\frac{1}{3}\epsilon_{ijkm}P^m\,.
\end{equation}

The $a_k$ and $b_k$ in the Lagrangian are free coupling parameters.
Due to the Bach-Lanczos identity only five of the six $b_k$'s are
independent. $a_0$ is the coupling parameter of the scalar curvature
$R:=R_{\mu \nu}{}^{\nu \mu}$. Note that (because of the assumed
quadratic Lagrangian, linear-in-field strength canonical momenta)
one obtains, as in the standard physics paradigm, 2nd order
equations for the potentials by varying the Lagrangian $L_g$ of the
PGT independently with respect to the frame and connection. It
should be remarked that these PGT equations are quite different from
the problematical 4th order type of equations obtained from
Riemannian geometry based Lagrangians of the form
$R+(R_{..}{}^{..}){}^2$ when varied with respect to the metric.

In the PGT, in addition to the dynamic metric represented by the
translational gauge potential (the orthonormal frame), the
rotational gauge potential (the connection) has some independent
dynamics. As in other gauge theories, it is usually convenient to
describe the dynamics of the connection (a non-covariant, gauge
dependent potential) in terms of a tensorial field strength.  In the
PGT case these modes can be described by the torsion tensor. As
mentioned in the introduction, the various PGT dynamic torsion modes
were first investigated via the linearized theory; it was shown that
the torsion decomposes into six modes with certain spins and parity:
$2^\pm,1^\pm,0^\pm$. Later investigations
\cite{HNZ96,CNY98,YN99,YN02} concluded that effects due to
non-linearities in the constraints could be expected to render all
of these cases physically unacceptable except for the two ``scalar
torsion'' modes: spin-0$^+$ and spin-0$^-$. These two dynamic scalar
torsion modes apparently are the only physically acceptable dynamic
PGT torsion modes.

%%%%%%%%%%%%%%%%%%%%%%%%%%%%%%%%%%%%%%%%%%%%%%
\subsection{simple spin-$0^+$ mode}
%%%%%%%%%%%%%%%%%%%%%%%%%%%%%%%%%%%%%%%%%%%%%%%%%%
Here we only investigate the {\it simple} spin-0$^+$ case, i.e.,
choosing $a_2=-2a_1$, $a_3=-a_1/2$ and taking all the $b_k$'s to
vanish except for $b_6=b\ne0$.  (For a detailed analysis of this
case please see \cite{YN99}.) Our gravitational Lagrangian density
%\cite{Lagrangian}
for this spin-0$^+$ mode is then
\begin{eqnarray}
  L_g &=& -\frac{a_0}{2}R +\frac{b}{24}R^2\nonumber\\
        &&+\frac{a_1}{8}(T_{\nu\sigma\mu}T^{\nu\sigma\mu}
         +2T_{\nu\sigma\mu}T^{\mu\sigma\nu}-4T_\mu
         T^\mu)\,,\label{Lg0+}
\end{eqnarray}
where $T_\mu:=T_{\mu\nu}{}^\nu$. The Hamiltonian analysis showed
that the number of degrees of freedom in  $L_g$ is three: the scalar
torsion mode and two helicity states of the usual massless graviton
(provided the scalar torsion mode is massive, i.e, $a_1\ne a_0$).
%Positivity of the kinetic energy density for the dynamic components requires
It is necessary to impose certain sign conditions on the parameters
(see \cite{HS80,SN80,BMNI83,NicI84,YN99,Blag02}):
\begin{equation}
a_1>0\,,\qquad b>0\,.
\end{equation}
There is a simple argument which accounts for the signs of these two
parameters.  In order to have {\em least action} the kinetic energy
contribution from any dynamic variable must be positive (for if such
a term were negative the action would have no lower bound, since we
could have an arbitrarily large time rate of change for a dynamic
variable). Consider that $b$ is the parameter associated with the
quadratic scalar curvature term $R^2$. With the help of
Eq.~(\ref{defR}), it can be seen that the scalar curvature includes
some time derivatives of one of our basic dynamic fields, the
connection components: $R^2=(e^i{}_\nu e^j{}_\mu
R_{ij}{}^{\mu\nu})=(2e^t{}_\nu e^j{}_\mu{\dot\Gamma}_j{}^{\mu
\nu}+\cdots)^2= 4(e^t{}_\nu e^j{}_\mu{\dot\Gamma}_j{}^{\mu
\nu})^2+\cdots\ge0$.
%, where $(e^t{}_0
%e^j{}_\mu{\dot\Gamma}_j{}^{\mu 0})^2$ is the principal part of the
%kinetic term of $R^2$. Since $(e^t{}_0
%e^j{}_\mu{\dot\Gamma}_j{}^{\mu 0})^2>0$, $b$ is supposed to be
%positive for a positive kinetic energy in the Lagrangian density
%(\ref{Lg0+}).
Hence the coefficient of this term in the action should be positive.
A similar argument based on (\ref{torsiondef}), taking into account
the chosen metric signature and the restricted form of the torsion
in our model as a consequence of the field equations,
Eq.~(\ref{restrictedT}) below, gives the sign of $a_1$.

Varying $L_g$ (\ref{Lg0+}) with respect to the potentials
$e_i{}^\mu, \Gamma_i{}^{\mu\nu}$ gives gives the specific second
order field equations of the general form (\ref{first},\ref{second})
for this mode.
 Assuming $S_{\mu\nu}{}^i=0$
(i.e., the source spin current is negligible) we investigate first
Eq.~(\ref{second}), obtaining for this mode the three decomposed
equations:
\begin{eqnarray}
  \nabla_\mu R&=&-\frac{2}{3}(R+\frac{6\mu}{b})T_\mu\,,\label{pgtfe1}\\
             0&=&-(R+\frac{6\mu}{b})P_\mu\,,\\
             0&=&-(R+\frac{6\mu}{b})Q_{\mu\nu\sigma}\,,
\end{eqnarray}
where $\mu:=a_1-a_0$ is the effective mass of the linearized $0^+$
mode. There is a special case with no dynamical scalar torsion if
$R=-6\mu/b$, a constant. We do not treat this exceptional degenerate
situation as an isolated case (it is considered below as a limit of
the generic case). Assuming that $R+6\mu/b\neq0$ generically leads
to
\begin{equation}
P_\mu=Q_{\mu\nu\sigma}=0\, .\label{vanish}
\end{equation}
Using these two constraints gives the restricted form of the
torsion:
\begin{equation}
  T_{ij}{}^\mu=\frac{2}{3}T_{[i}e_{j]}{}^\mu.\label{restrictedT}
\end{equation}
Substituting into  Eq.~(\ref{first}) with our specific parameter
choices gives the restricted field equation
\begin{eqnarray}
  \nabla_jH_\mu{}^{ij}-E_\mu{}^i&=&e\Bigl\{ {2a_1\over3}[e^i{}_\nu\nabla_\mu T^\nu
     -e^i{}_\mu\overline{\nabla}_jT^j] \nonumber\\
    &&-e^i{}_\mu[ - {a_0\over2}R+ {b\over24}R^2
  -\frac{a_1}{3}T_iT^i] \nonumber\\
   &&+ R_\mu{}^i[{b\over6}R-a_0]\Bigr\}={\cal T}_\mu{}^i.
  \label{pgtfe2}
\end{eqnarray}
Now we have a complete set of the field equations, (\ref{pgtfe1})
and (\ref{pgtfe2}) along with (\ref{vanish}). In \cite{HNZ96} it was
argued that this system had a well posed initial value problem.
However that may be, the two main field equations are rather
complicated. They really look nothing like the familiar,
well-analyzed equations of GR. To help understand the significance
of these new relations, and to use our previous experience, we will
do a translation of (\ref{pgtfe1},\ref{pgtfe2}) into a certain
effective Riemannian form---transcribing from quantities expressed
in terms of the orthonormal tetrad $e_j{}^\mu$ and connection
$\Gamma_i{}^{\mu\nu}$ into the ones expressed in terms of the metric
$g_{jk}$ and torsion $T_{ij}{}^k$. Then we can compare the result
with the more familiar field equations in general relativity.
%%%%%%%%%%%%%%%%%%%%%%%%%%%%%%%%%%%%%%%%%%%%%%%%%%%%%
\subsection{Translation}
%%%%%%%%%%%%%%%%%%%%%%%%%%%%%%%%%%%%%%%%%%%%%%%%%%%%%
As is well-known, the PGT affine connection can be represented in
the form
\begin{equation}
  \Gamma_{ij}{}^k=\overline{\Gamma}_{ij}{}^k+\frac{1}{2}(T_{ij}{}^k+T^k{}_{ij}
  +T^k{}_{ji})\,,
\end{equation}
where $\overline{\Gamma}_{ij}{}^k$ is the Levi-Civita connection,
\begin{equation}
  \overline{\Gamma}_{ij}{}^k=\frac{1}{2}g^{km}(g_{mj,i}+g_{mi,j}-g_{ij,m})\,,
\end{equation}
and $T_{ij}{}^k$ is the torsion. Accordingly the affine Ricci
curvature and scalar curvature can be represented as
\begin{eqnarray}
   R_{ij} &=& \overline{R}_{ij} + \overline{\nabla}_jT_i+\frac{1}{2}
   (\overline{\nabla}_k - T_k)(T_{ji}{}^k+T^k{}_{ij}+T^k{}_{ji})\nonumber\\
   &&+\frac{1}{4}(T_{kmi}T^{km}{}_j+2T_{jkm}T^{mk}{}_i)\,,\\
   R&=&\overline{R} + 2\overline{\nabla}_iT^i+\frac{1}{4}(T_{ijk}T^{ijk}
    +2T_{ijk}T^{kji}-4T_iT^i)\,,\nonumber\\ &&
\end{eqnarray}
where $\overline{R}_{ij}$ and $\overline{R}$ are the Riemannian
Ricci curvature and scalar curvature, respectively, and
$\overline\nabla$ is the covariant derivative with the connection
$\overline{\Gamma}_{ij}{}^k$.

For the case of interest here the torsion tensor has the restricted
form (\ref{restrictedT}).
%\begin{equation}
%   T_{ijk} = \frac{2}{3}T_{[i}g_{j]k}\,,
%\end{equation}
%where the vector $T_i$ is the trace of the torsion.
 Consequently the
affine Ricci curvature and scalar curvature become
\begin{eqnarray}
   R_{ij}&=&\overline{R}_{ij}+\frac{1}{3}(2\overline{\nabla}_jT_i+g_{ij}
   \overline{\nabla}_kT^k)\nonumber\\
   &&+\frac{2}{9}(T_iT_j-g_{ij}T_kT^k)\,,\label{aRicvsrRic}\\
   R&=&\overline{R} + 2\overline{\nabla}_iT^i-\frac{2}{3}T_iT^i\,.\label{aRvsrR}
\end{eqnarray}
Applying this translation {\em selectively} \cite{highord} in
Eqs.~(\ref{pgtfe1}) and (\ref{pgtfe2}) gives an alternate form of
the field equations:
\begin{eqnarray}
   &&\overline{\nabla}_iR +\frac{2}{3}(R+\frac{6\mu}{b})T_i=0\,,\label{gradR}\\
   &&a_0(\overline{R}_{ij}-\frac{1}{2}g_{ij}\overline{R})-\frac{b}{6}R
        (R_{(ij)}-\frac{1}{4}g_{ij}R)\nonumber\\
   &&\qquad\qquad-\frac{2\mu}{3}(\overline{\nabla}_{(i}T_{j)}
   -g_{ij}\overline{\nabla}_kT^k) \nonumber\\
   &&\qquad\qquad-\frac{\mu}{9}(2T_iT_j+g_{ij}T_kT^k) =-{\cal T}_{ij}\,,\label{fe1}
\end{eqnarray}
while contracting Eq.~(\ref{fe1}) with the help of
Eq.~(\ref{aRvsrR}) yields
\begin{equation}\label{RRbar}
 a_1\overline{R}-\mu R={\cal T}\,.
\end{equation}

Note that the relation (\ref{fe1}) can be re-written into the form
of Einstein's equation:
\begin{equation}\label{effGR}
   a_0(\overline{R}_{ij}-\frac{1}{2}g_{ij}\overline{R})
  =-\tau_{ij}:=-({\cal T}_{ij}+\widetilde{\cal T}_{ij})\,,
\end{equation}
where ${\cal T}_{ij}$ is the source energy-momentum tensor and the
contribution of the scalar torsion mode to the effective total
energy-momentum tensor $\tau_{ij}$ is
\begin{eqnarray}
   \widetilde{\cal T}_{ij}&=&-\frac{2\mu}{3}(\overline{\nabla}_{(i}T_{j)}
  -g_{ij}\overline{\nabla}_kT^k)-\frac{\mu}{9}(2T_iT_j+g_{ij}T_kT^k)\nonumber\\
    &&-\frac{b_6}{6}R(R_{(ij)}-\frac{1}{4}g_{ij}R)\,.\label{torT}
\end{eqnarray}
However, it should be kept in mind that ${\widetilde{\cal T}}_{ij}$
is only an effective quantity. Using this effective quantity allows
us to use some of the insight we have obtained from our experience
with GR. Thus we can regard the contribution of $\widetilde{\cal
T}_{ij}$ to the rhs of (\ref{effGR}) as something like that of an
exotic field. This hybrid form is practical for our needs here, even
though it is not really a proper fundamental physical description
(one way to see this is to note that $\widetilde{\cal T}_{ij}$
cannot be obtained as the Hilbert energy-momentum density of some
effective source Lagrangian).

%(Note by the way that expanding all the affine curvature terms in
%(\ref{gradR},\ref{fe1}) using (\ref{aRicvsrRic},\ref{aRvsrR}) would
%lead to unpleasant equations with 3rd derivatives of the metric and
%quadratic terms in the Riemannian curvature---not at all like GR.)

 Eqs.~(\ref{effGR},\ref{torT}) do allow us to
appreciate some of the similarities and differences between this
model and other accelerating universe models. However, to properly
understand this model, one should consider the torsion dynamics
geometrically rather than trying to regard it as just another field
in a Riemannian spacetime obeying Einstein's equations.

It is remarkable that the torsion vector---as a consequence of the
field equation (\ref{gradR})---turns out to be the gradient of a
scalar function. In fact we can identify the function as $-(3/2)\ln
(R+6\mu/b)$. However (especially given its geometric nature) we do
not see any way to introduce this scalar potential directly into the
Lagrangian as a fundamental field. Were that possible one could then
directly compare features of our scalar torsion model with the
various scalar field dark energy models.  But as far as we can see,
notwithstanding a few similarities, our model is really not much
like those scalar field models.

%Here we show eqns (\ref{effGR}-\ref{torT}) in order to let people
%appreciate the similarities and differences between this model and
%other cosmological models. To understand this model correctly, we
%still need to consider metric and torsion in a Rieman-Cartan
%spacetime instead of considering only metric in a Riemannian
%spacetime.

%%%%%%%%%%%%%%%%%%%%%%%%%%%%%%%%%%%%%%%%%%%%%%%%%%%%%
\section{Field Equations For Torsion Cosmology}
%%%%%%%%%%%%%%%%%%%%%%%%%%%%%%%%%%%%%%%%%%%%%%%%%%%%%
For cosmology, assuming homogeneous and isotropic leads to the
Friedmann-Robertson-Walker metric
\begin{eqnarray}
 {\rm d}s^2 = -{\rm d}t^2+a^2(t)\Bigl[\frac{{\rm d}r^2}{1-kr^2}
 +r^2({\rm d}\theta^2+\sin^2\theta{\rm d}\phi^2)\Bigr],
\end{eqnarray}
where $a(t)$ is the expansion factor, and $k$ is the curvature
index. Here, to see the effects we are interested in as well as
match the observations, it is sufficient to consider only the
simplest case: the flat universe with $k=0$. This yields the
non-vanishing (Riemannian) Ricci and scalar curvature:
\begin{eqnarray}
  &&\overline{R}_t{}^t = 3\frac{\ddot{a}}{a} = 3(\dot{H} + H^2)\,,\\
  &&\overline{R}_r{}^r = \overline{R}_\theta{}^\theta =\overline{R}_\phi{}^\phi
                  =\frac{\ddot{a}}{a} + 2\frac{\dot{a}^2}{a^2}
                  = \dot{H} + 3H^2\,,\\
  &&\overline{R}=6(\frac{\ddot{a}}{a}+\frac{\dot{a}^2}{a^2})=6(\dot{H} + 2H^2)\,,
\end{eqnarray}
where $H:=\dot a/a$.  The torsion $T_i$ should also be only time
dependent, i.e., $T_i$=$T_i(t)$. So from (\ref{gradR}) the spatial
parts of $T_i$ vanish. Letting $T_t(t)$=$\Phi(t)$ we have
\begin{equation}
  \dot{R} =-\frac{2}{3}\left(R+\frac{6\mu}{b}\right)\Phi.
\end{equation}
Integrating this equation leads to
\begin{eqnarray}
   R&=&6(\dot{H} + 2H^2 -H\Phi) -2\dot{\Phi} +\frac{2}{3}\Phi^2\nonumber\\
    &=&-\frac{6\mu}{b} +\left(R(t_0)+\frac{6\mu}{b}\right)
       \exp\left(-\frac{2}{3}\int^t_{t_0}\Phi dt'\right).\label{afftrR}
\end{eqnarray}
From the field equations we can finally give the necessary equations
to integrate:
\begin{eqnarray}
      \dot{a}&=&aH\,,\label{dta}\\
      \dot{H}&=&\frac{\mu}{6a_1}R+\frac{1}{6a_1}{\cal T}-2H^2\,,\label{dtH}\\
      \dot{\Phi}&=&-\frac{a_0}{2a_1}R+\frac{1}{2a_1}{\cal T}-3H\Phi
                   +\frac{1}{3}\Phi^2\,,\label{dtphi}\\
      \dot{R}&=&-\frac{2}{3}\left(R+\frac{6\mu}{b}\right)\Phi\,,\label{dtR}
\end{eqnarray}
where
\begin{eqnarray}
  &&\frac{b}{18}(R+\frac{6\mu}{b})(3H-\Phi)^2-\frac{b}{24}R^2-3a_1H^2
   ={\cal T}_{tt}=\rho\,,\nonumber\\
   &&\label{fieldrho}\\
  &&{\cal T} = g^{ij}{\cal T}_{ij} = 3p-\rho\,,\\
   &&p=w\rho\,.
\end{eqnarray}
Here we consider only the matter-dominated era, where the pressure
$p$ is negligible.

For the effective energy-momentum tensor contribution from the
scalar torsion mode $\widetilde{\cal T}_{ij}$, the explicit
expression is:
\begin{eqnarray}
  \widetilde{\cal T}_t{}^t&=&\!\!-3\mu H^2\!+\frac{b}{18}(R+\frac{6\mu}{b})
  (3H-\Phi)^2-\frac{b}{24}R^2,\label{torho}\\
   \widetilde{\cal T}_r{}^r&=&\widetilde{\cal T}_\theta{}^\theta
   =\widetilde{\cal T}_\phi{}^\phi
     ={1\over 3}[\mu(R-\overline{R})-\widetilde{\cal T}_t{}^t]\,,\label{torpre}
\end{eqnarray}
and the off-diagonal terms vanish.

We define $\rho_{\rm eff}\equiv\rho+\rho_{\rm T}=-3a_0H^2$, where
$\rho_{\rm T}\equiv\widetilde{\cal T}_{tt}$ is the torsion-induced
mass density. $\rho_{\rm eff}$ means the effective mass density
which is deduced from general relativity. $p_{\rm
T}\equiv\widetilde{\cal T}_r{}^r$ is an effective pressure due to
contributions induced by the dynamic torsion.
%%%%%%%%%%%%%%%%%%%%%%%%%%%%%%%%%%%%%%%%%%%%%%%%%%%%%%%%%%%%%%%%%%
\section{A Preliminary Analysis of the Equations}
%%%%%%%%%%%%%%%%%%%%%%%%%%%%%%%%%%%%%%%%%%%%%%%%%%%%%%%%%%%%%%%%%%
Equations (\ref{dta}--\ref{dtR}) are the main equations for the
integrations to evolve the system.  Regarding the parameters in the
field equations, the Newtonian limit requires $a_0\equiv-(8\pi
G)^{-1}$. We take $a_1>0$ and $b>0$ to satisfy the energy positivity
requirement \cite{YN99}. Moreover the no tachyon condition for the
scalar torsion is then also satisfied: $\mu=a_1-a_0>(8\pi
G)^{-1}>0$.

Before the detailed results are shown, we briefly analyze the
equations to obtain some insight about their behavior. Let us first
study the behavior of the affine scalar curvature $R$. The second
derivative of $R$ with respect to time can be obtained by operating
a time derivative on Eq.~(\ref{dtR}) and using Eq.~(\ref{dtphi}):
\begin{equation}\label{dttR}
  \ddot{R}=-\frac{2}{3}\dot R\Phi-\frac{2}{3}(R+\frac{6\mu}{b})\dot\Phi
  \approx\frac{2a_0\mu}{a_1b}R\,,
\end{equation}
here we assumed that all the variables, i.e., $H$, $\Phi$, $R$ are
much smaller than the coefficient of the leading order term, i.e.,
$2a_0\mu/a_1b$. For ${\cal T}$, which appears on the rhs of
(\ref{dtphi}), we know it consists at least of quadratic terms of
$H$, $R$, and $\Phi$ from (\ref{fieldrho}), so it should be smaller
than the other variables. This shows that the coefficient of $R$ on
the right hand side of (\ref{dttR}) is negative: $2a_0\mu/a_1b<0$.
From this analysis we find that the late-time behavior of $R$ will
be essentially oscillating with the period
\begin{equation}\label{anaT}
 T=2\pi\sqrt{-\frac{a_1b}{2a_0\mu}}\,.
\end{equation}
By a similar argument, it is easy to infer that $\Phi$ has the same
periodical behavior.

Next we direct our attention to the behavior of the expansion factor
$a$. The acceleration of the expansion factor $a$ can be obtained by
combining
Eqs.~(\ref{RRbar},\ref{dta},\ref{dtH},\ref{torho},\ref{torpre}):
\begin{equation}\label{ddota}
\ddot{a}=\frac{\mu R+{\cal T}}{6a_1}a-\frac{\dot{a}^2}{a}
        =\frac{3p_{\rm T}+\rho_{\rm eff}}{6a_0}a\,.
\end{equation}
From this the relation between the acceleration of the expansion
$\ddot a$ and the quantity $3p_{\rm T}+\rho_{\rm eff}$ can be
clearly seen. Since $a_0<0$, it shows that $\ddot{a}>0$ as long as
$3p_{\rm T}+\rho_{\rm eff}<0$, and vice versa. We will discuss this
 relation and its demonstration in the next section.

The period of $a$ and $H$, if they exist, should be same as that of
$\Phi$ and $R$. Because the variables are all highly coupled to each
other to form an equation set, there should generically exist a
common period in the solution. This point will be demonstrated in
the later numerical analysis.

We need to look into the scaling features of this model
before we can obtain the sort of results we seek on a cosmological
scale. In terms of fundamental units we can scale the variables and
the parameters as
\begin{eqnarray}
t\rightarrow t/\ell,\,\, a\rightarrow a,\,\, H\rightarrow\ell H,
\,\, \Phi\rightarrow\ell\Phi,\,\, R\rightarrow\ell^2R,&&\nonumber\\
a_0\rightarrow\ell^2a_0,\,\, a_1\rightarrow
\ell^2a_1,\,\, \mu\rightarrow\ell^2\mu,\,\, b\rightarrow b,&&\label{scale1}
\end{eqnarray}
where $\ell\equiv\sqrt{8\pi G}$. So the variables and the scaled
parameters $a_0$, $a_1$, and $b$ all become dimensionless, and
$a_0=-1$. Furthermore, Eqs.~(\ref{dta}--\ref{dtR}) remain unchanged
under such a scaling. However, as we are interested in the
cosmological scale, it is practical to use another scaling to turn
the numerical values of the scaled variables ``gentler'' (i.e., not
stiff) from the numerical integration. In order to achieve this
goal, let us introduce a dimensionless constant $T_0$, which
represents the magnitude of the Hubble time
($T_0=H^{-1}_0\doteq4.41504\times 10^{17}$ seconds.) Then the
scaling is
\begin{eqnarray}
t\rightarrow T_0t,\,\, a\rightarrow a,\,\, H\rightarrow H/T_0,\,\,
\Phi\rightarrow\Phi/T_0,&&R\rightarrow R/T_0^2,\nonumber\\
a_0\rightarrow a_0,\,\, a_1\rightarrow a_1,\,\, \mu\rightarrow\mu,\,\,
b\rightarrow T^2_0 b,&&\label{scale2}
\end{eqnarray}
With this scaling, all the field equations are kept unchanged while
the period $T\rightarrow T_0T$.
%%%%%%%%%%%%%%%%%%%%%%%%%%%%%%%%%%%%%%%%%%%%%%%%%%%%%%%%%%%%%%%%%%
\subsection{parameter choice with constant scalar curvature}
%%%%%%%%%%%%%%%%%%%%%%%%%%%%%%%%%%%%%%%%%%%%%%%%%%%%%%%%%%%%%%%%%%
\label{anaA} Equation (\ref{dtR}) is of special interest among the
field equations because of the existence of a constant value
$6\mu/b$. It shows that the scalar affine curvature remains a
constant $R=-6\mu/b$ forever as long as its initial data has this
special critical value. It is tempting to see how the system evolves
with $R=-6\mu/b$ initially.

As mentioned in Sec.~\ref{secpgt}, the positivity of the kinetic
energy in the Hamiltonian analysis of the spin-0$^+$ case requires
$b>0$, $a_1>0$, thus $\mu>0$ since $a_0<0$. With such an assumption,
the scalar affine curvature should not have the value $R=-6\mu/b$
initially since this initial choice will require the matter density
$\rho$ to be negative from Eq.~(\ref{fieldrho}). Such a  choice
violates the assumption of energy positivity.

However, if we tentatively relax the parameter requirement for
positive kinetic energy, i.e., allowing $a_1=-{\bar a}_1<0$ such
that $\mu=-m<0$, the scenario will turn out to be quite intriguing.
Under such a new parameter requirement, if we set initially the
scalar affine curvature
\begin{equation}
R=-\frac{6\mu}{b}=\frac{6m}{b}>0\,,
\end{equation}
then $R$ will remain at this constant value for all the time. From
Eqs.~(\ref{fieldrho}) and (\ref{torho}), we can derive
\begin{equation}
-3a_0H^2=\rho+\rho_{\rm T}>0\,,
\end{equation}
where
\begin{eqnarray}
\rho&=&3{\bar a}_1H^2-\frac{3}{2}\frac{m^2}{b},\\
\rho_{\rm T}&=&\frac{3}{2}\frac{m^2}{b}-3mH^2.
\end{eqnarray}
Here the matter density will be positive as long as the parameters
are chosen suitably, such that ${\bar a}_1H^2-m^2/2b>0$. And the
more interesting point is that the torsion-induced mass density
$\rho_{\rm T}$ could ``act like a dark energy'' if the suitable
parameter values are chosen. We can simplify the field equation
(\ref{dtH}) to
\begin{equation}
\dot H=\frac{3}{4}\frac{m^2}{{\bar a}_1b}-\frac{3}{2}H^2,\label{dtHconR}
\end{equation}
and it leads to
\begin{equation}
\ddot a=\frac{1}{2}\left(\frac{3}{2}\frac{m^2}{{\bar
a}_1b}-H^2\right)a\,.
\end{equation}
Combining with Eq.~(\ref{ddota}), $3m^2/2{\bar a}_1b>H^2$ as long as
$3p_{\rm T}+\rho_{\rm eff}<0$. By solving Eq.~(\ref{dtHconR}), the
solution will show that
\begin{equation}
H\rightarrow\frac{m}{\sqrt{2{\bar a}_1b}}\quad{\rm for}\quad t\rightarrow\infty.
\end{equation}
By comparing this to a universe with a cosmological constant
$\Lambda$ where the Hubble function $H$ approaches to
$\sqrt{\Lambda/3}$ as $t\rightarrow\infty$, we can see how to choose
 suitable values such that the cosmological constant $\Lambda$ and
thus the dark energy can be mimicked in this torsion cosmological
model with a constant affine scalar curvature. We will demonstrate
numerically the behavior of this case in the next section.

\begin{figure*}[thbp]
\begin{tabular}{rl}
\includegraphics[width=8.8cm]{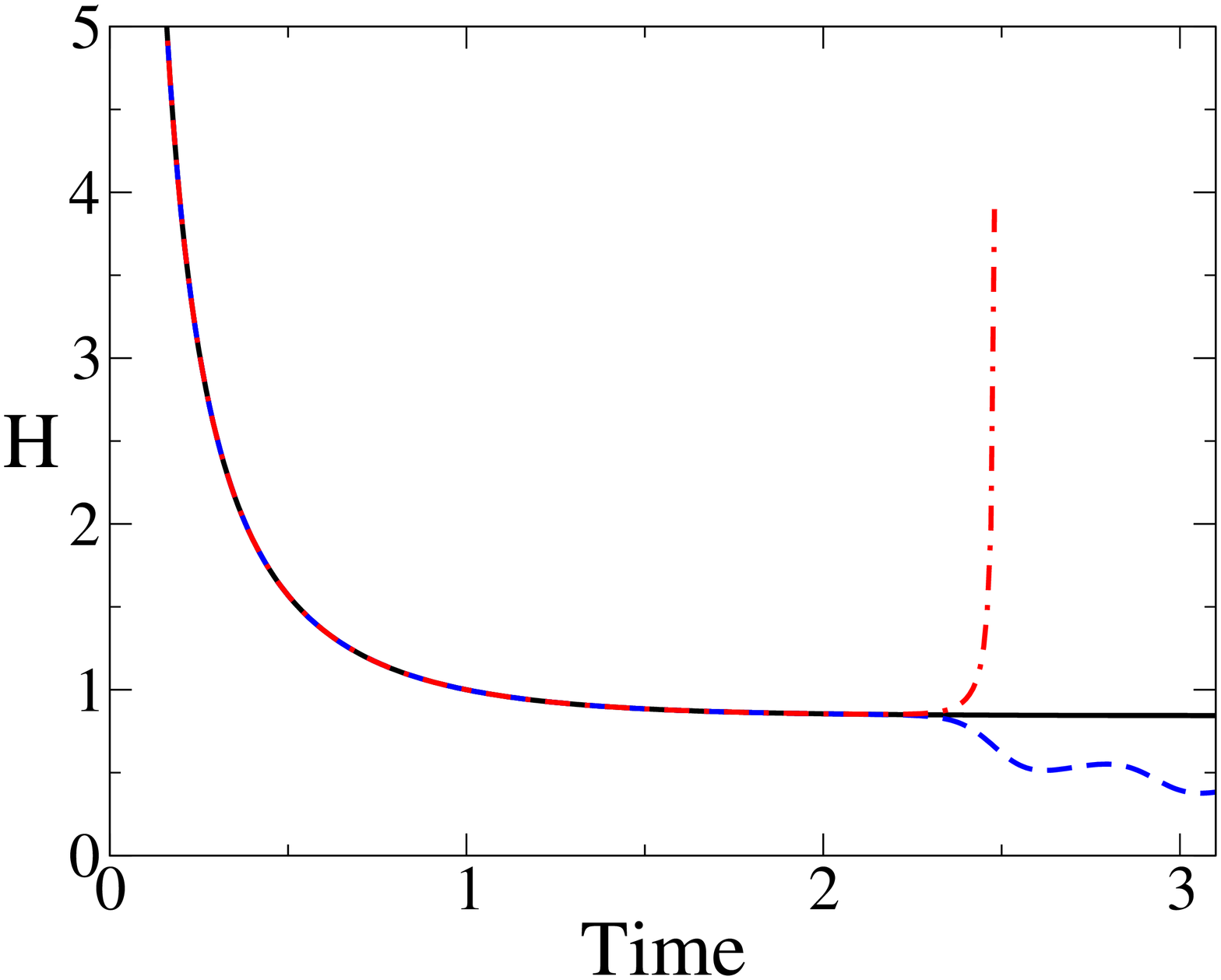}&
\includegraphics[width=8.8cm]{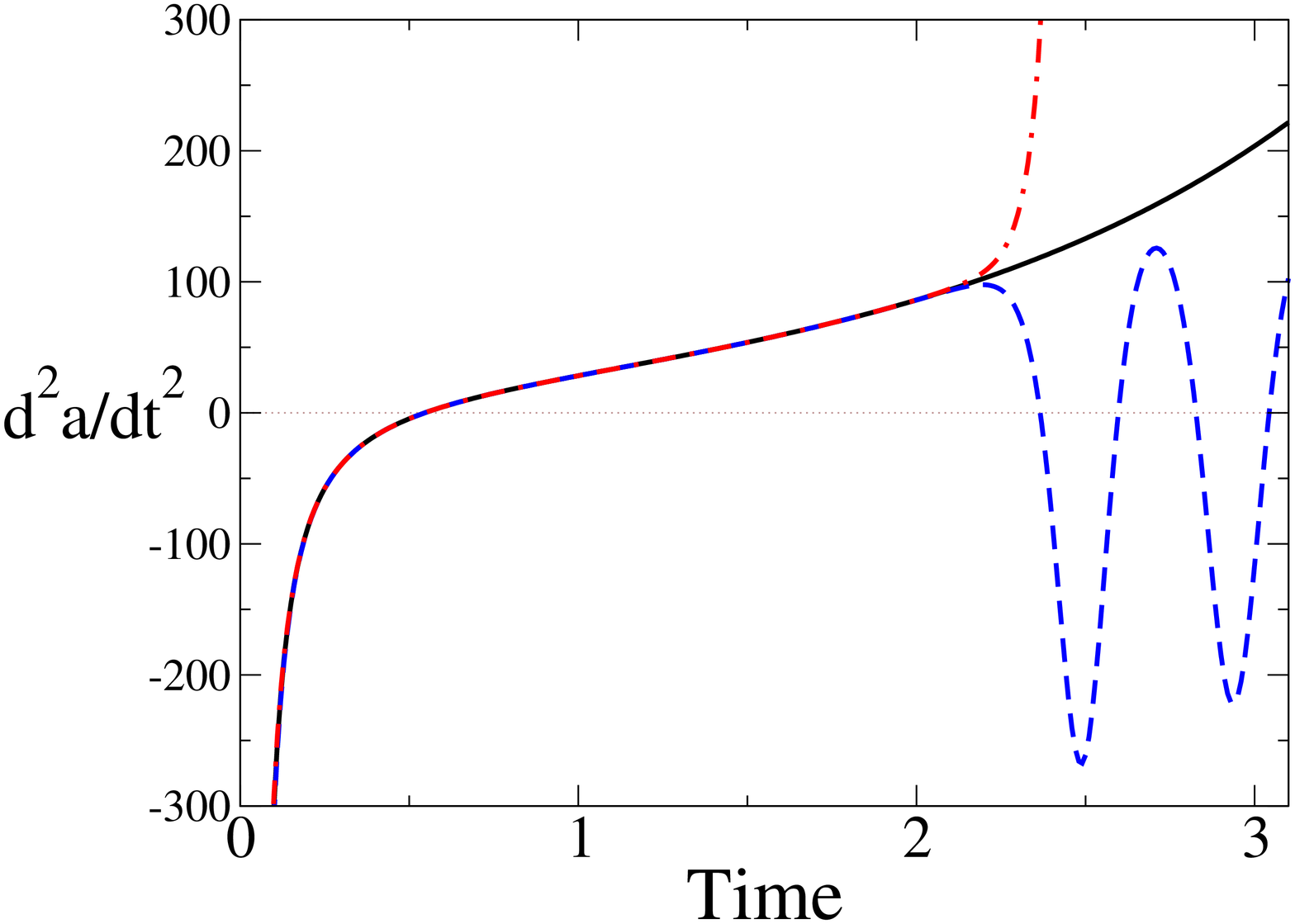} \\
\includegraphics[width=8.8cm]{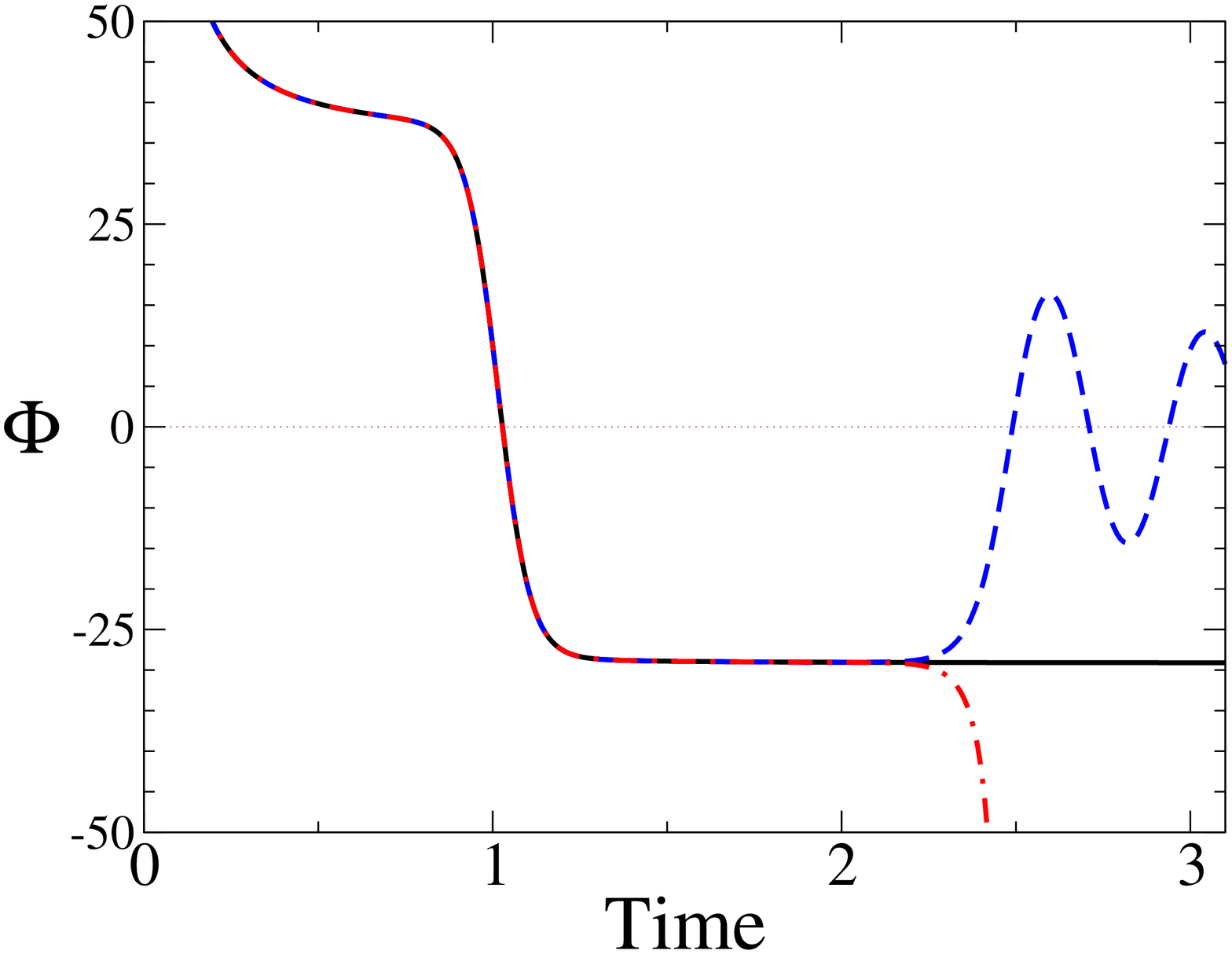}&
\includegraphics[width=8.8cm]{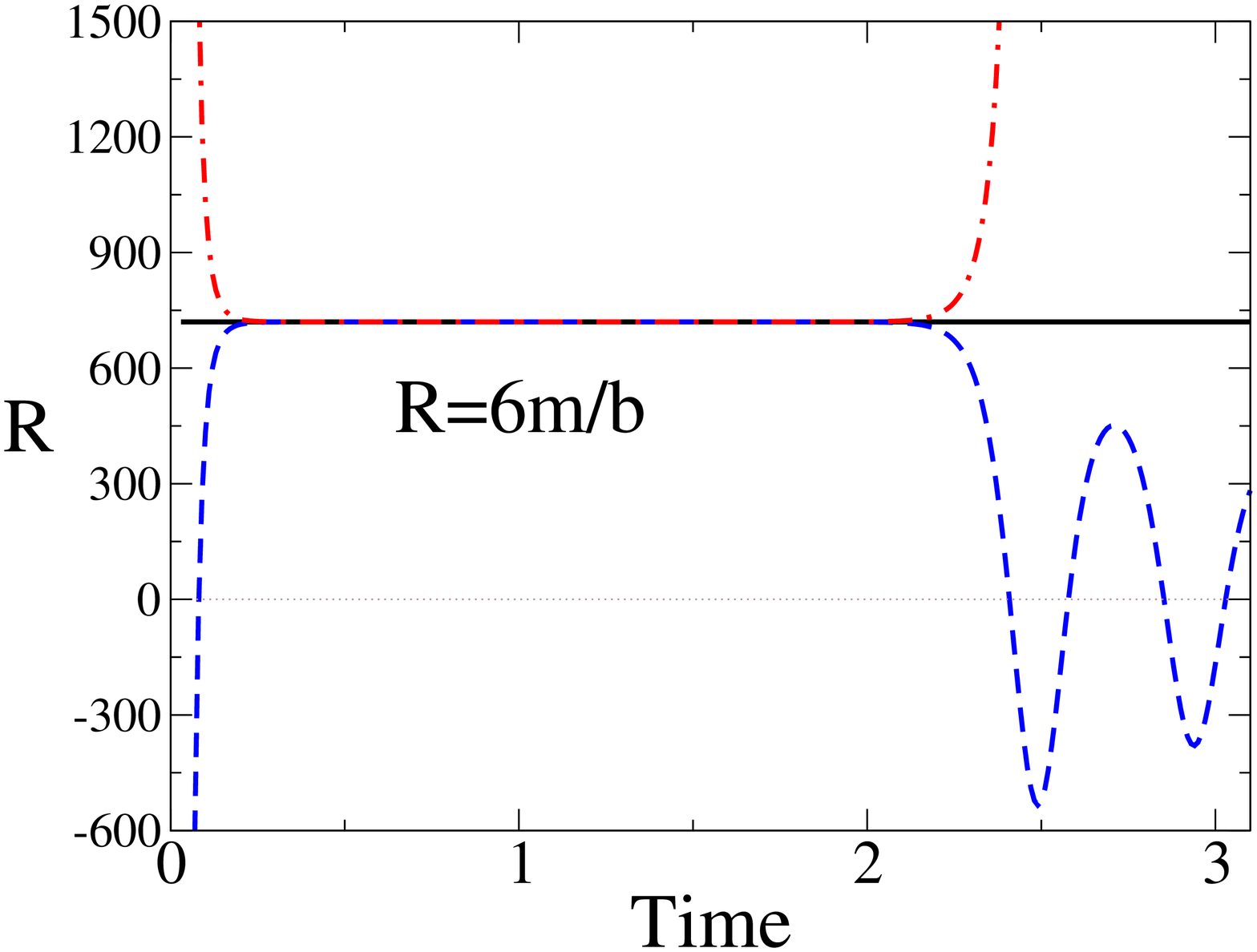}
\end{tabular}
\caption{Evolution of the Hubble function, $H$, the 2nd time derivative of
the expansion factor, $\ddot a$, the temporal component of the torsion,
$\Phi$, and the affine scalar curvature, $R$, as functions of time with the
parameter choice and the initial data in Case I.
The (black) solid lines represent the result of $R(t=0)=6m/b$,
the (blue) dashed lines represent the result of $R(t=1)=6m/b-10^{-8}$,
and the (red) dot-dashed lines represent the result of $R(t=1)=6m/b+10^{-8}$
while all the other initial choices are fixed.}
\label{conR}
\end{figure*}

\begin{figure*}[thbp]
\begin{tabular}{rl}
\includegraphics[width=8.8cm]{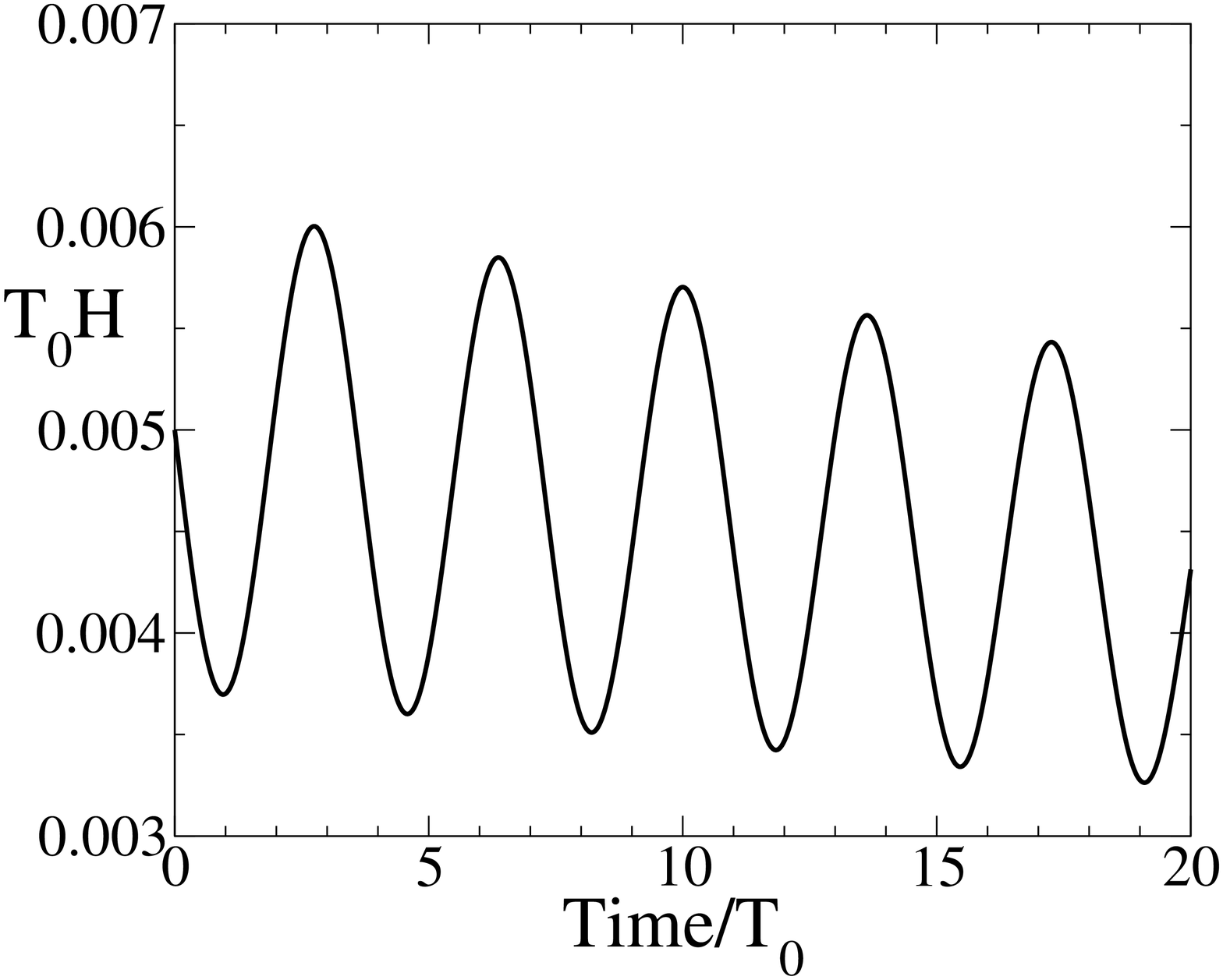}&
\includegraphics[width=8.8cm]{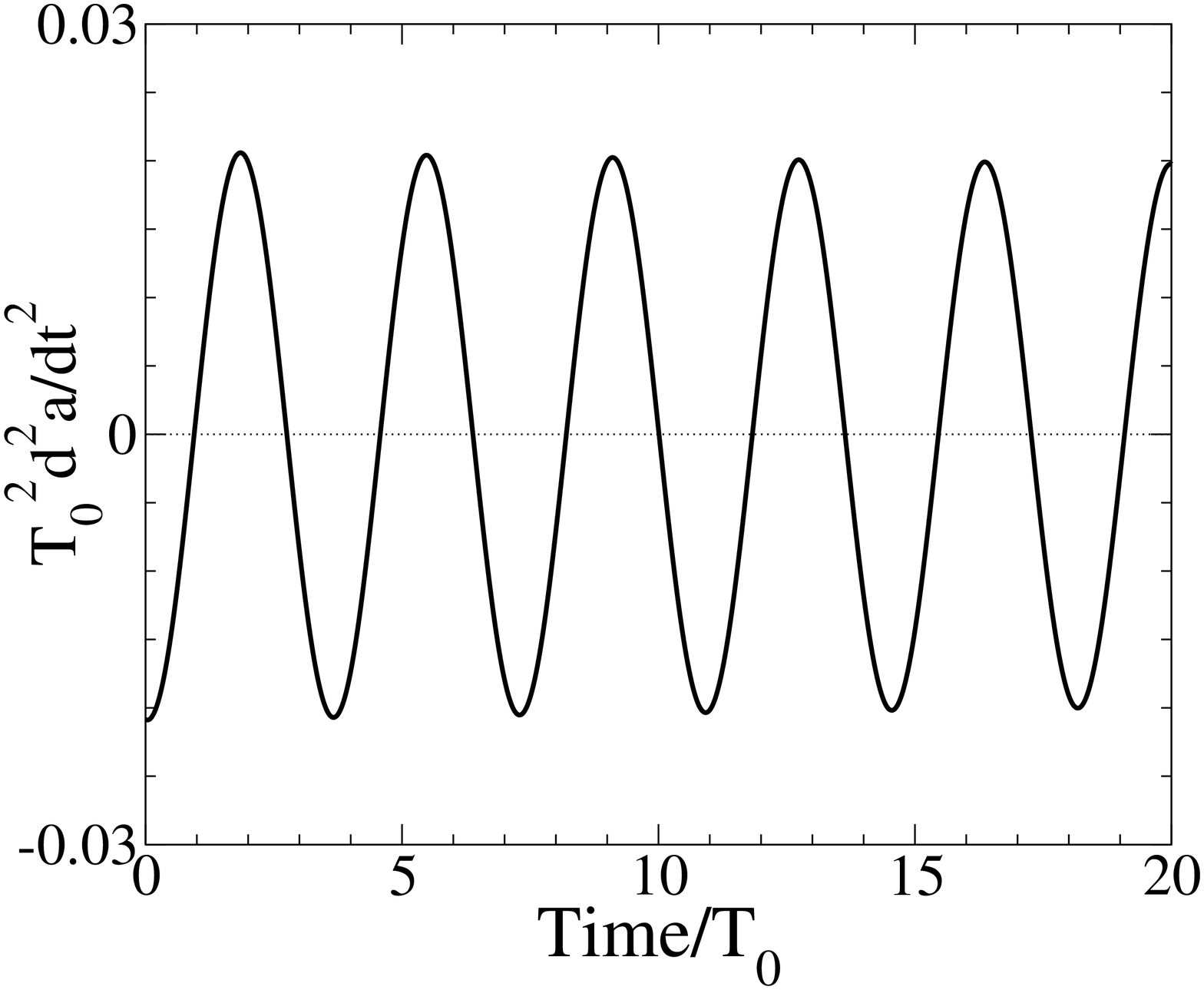} \\
\includegraphics[width=8.8cm]{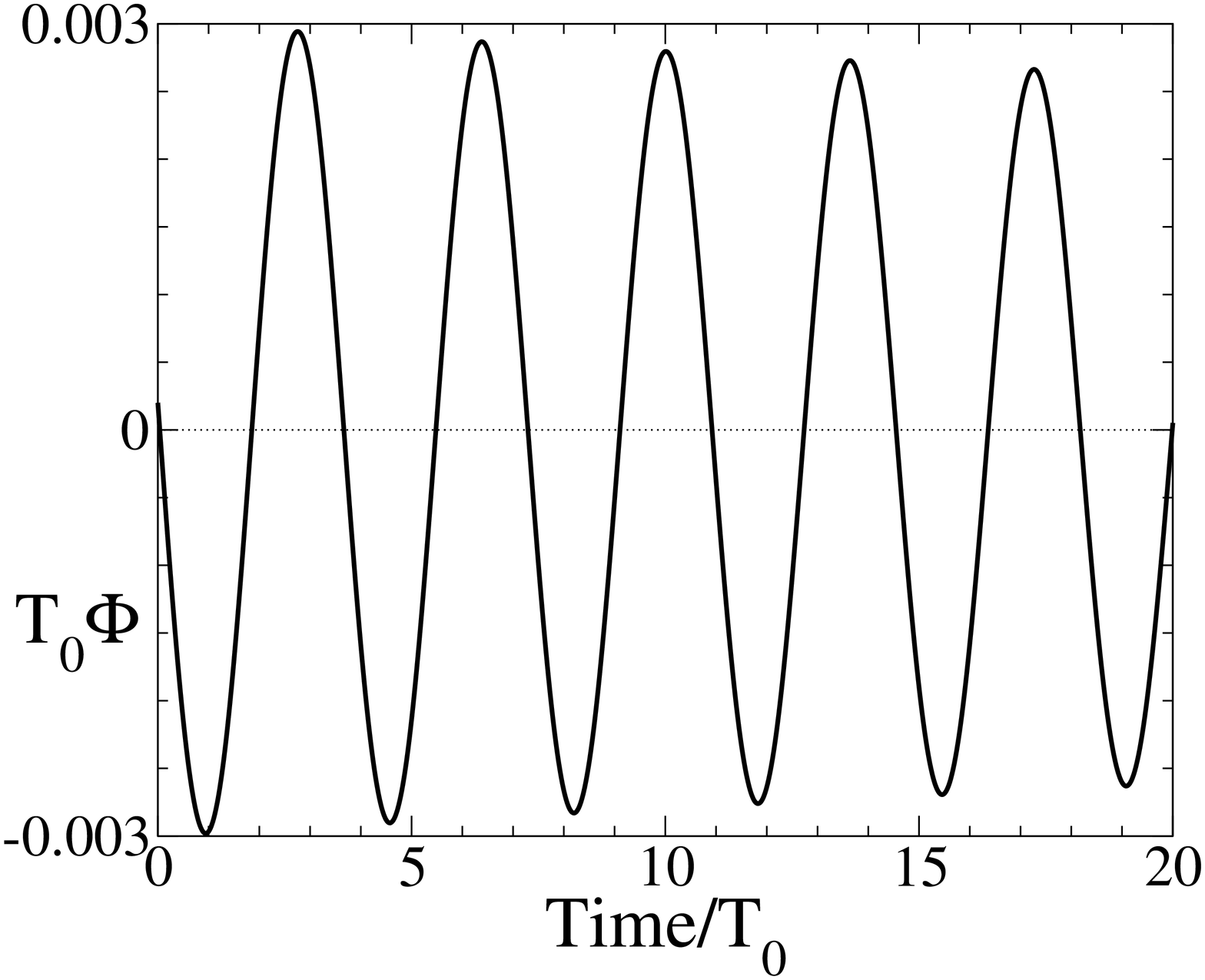}&
\includegraphics[width=8.8cm]{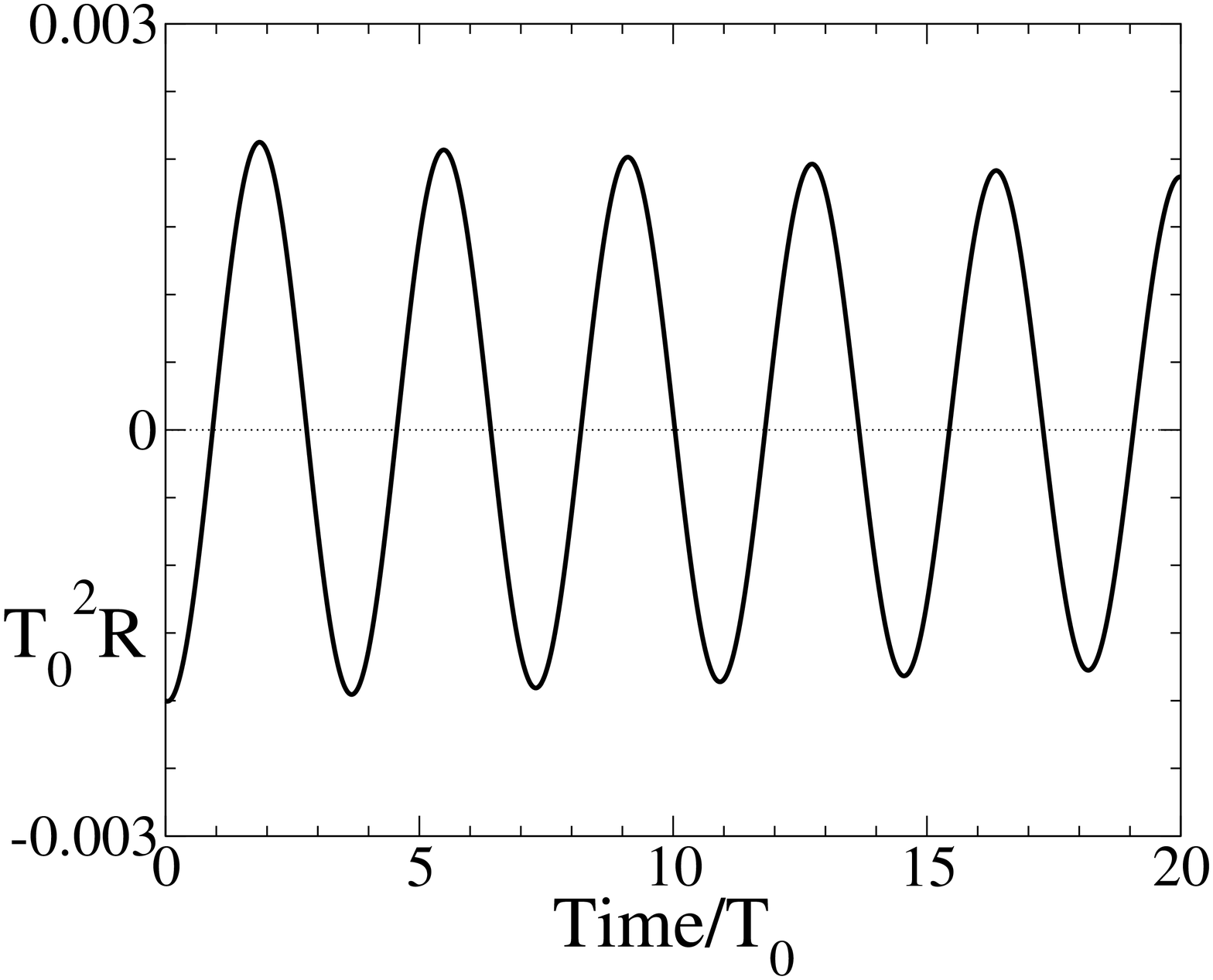}
\end{tabular}
\caption{Evolution of the Hubble function, the 2nd time derivative
of the expansion factor, the temporal component of the torsion, and
the affine scalar curvature as functions of time with the parameter
choice and the initial data in Case II.} \label{oscR1}
\end{figure*}

%%%%%%%%%%%%%%%%%%%%%%%%%%%%%%%%%%%%%%%%%%%%%%%%%%%%%%%%%%%%%%%%%%%%%%
\section{Numerical Demonstration}
%%%%%%%%%%%%%%%%%%%%%%%%%%%%%%%%%%%%%%%%%%%%%%%%%%%%%%%%%%%%%%%%%%%%%%
In this section we would like to demonstrate two points:  (1) The
degenerate case $R=-6\mu/b$ with the relaxed parameter choice
mentioned in subsection \ref{anaA}, i.e., $a_1<0$ and $\mu<0$
instead of the normal choice $a_1>0$ and $\mu>0$. Although such a
choice is against the positivity of kinetic energy, we would like to
explore this scenario a little bit more, since it could mimic the
cosmological constant and the other cosmological models with a
negative kinetic energy \cite{NKE}. We can see that the torsion in
the system becomes kinetic instead of being dynamic, and the
expansion is accelerating at late time; (2) In generic cases, i.e.,
$R+6\mu/b\ne0$, with the proper parameter choice (i.e., $a_1>0$ and
$\mu>0$), the torsion in the system is dynamic, and its functional
pattern has a periodic feature, i.e., it could be accelerating for a
while, and then be followed by a period of deceleration with the
pattern repeating. With suitable adjustments of the parameters and
the initial values of the fields involved, it is possible to change
the period of the dynamic fields as well as their amplitudes.
Furthermore, in the model, with some choices of the parameters and
the initial values of the fields, it is possible to mimic the main
apparent dynamic features of the Universe, i.e., the value of the
Hubble function is the current Hubble constant in an accelerating
universe after a period of time on the order of the Hubble time. In
such a case, this model will describe an oscillating universe with a
period on the order of magnitude of the Hubble time. This allows us
to constrain the parameters and/or the value of the torsion field by
comparing the observed data with the result from this model.

The 4th-order Runge-Kutta method is applied for the integration of
the field Eqs.~(\ref{dta}--\ref{dtR}). The Universe is assumed to be
matter-dominated, thus ${\cal T}\approx-\rho$. The mass density
$\rho$ is determined from the fields via Eq.~(\ref{fieldrho}). The
fields and the parameters are scaled with Eq.~(\ref{scale1}) and
Eq.~(\ref{scale2}) to be dimensionless, and to achieve a realistic
cosmology.
%%%%%%%%%%%%%%%%%%%%%%%%%%%%%%%%%%%%%%%%%%%%%%%%%%%%
\subsection{Case I: constant $R$ case}
%%%%%%%%%%%%%%%%%%%%%%%%%%%%%%%%%%%%%%%%%%%%%%%%%%%%
In this case, the initial values of the fields are as follows:
\[
  a(t_0)=50,\quad H(t_0)=1,\quad \Phi(t_0)=10,\quad R(t_0)=\frac{6m}{b},
\]
and the parameters are taken to be
\[
  a_0=-1,\quad b=10^{-4},
\]
where $t_0$ is the initial time: $t_0=1$, the present time of our
universe. Under this setting,
\begin{eqnarray}
\rho&=&-3a_0H^2+3mH^2-\frac{3m^2}{2b}\nonumber\\
 &=&3+3m-1.5\times10^4m^2.
\end{eqnarray}
In order that the mass density in the current universe is about
$\rho\approx 30\%$, the parameter $m$ is chosen as
\[
m=0.012\,.
\]
This shows that $H\rightarrow m/\sqrt{2{\bar a}_1b}\approx 0.84$.
The detailed result is shown in Fig.~\ref{conR}, where the evolved
values of $H$, $\ddot a$, $\Phi$, and $R$ are plotted as the (black)
solid curve in different panels.

It is obvious in the bottom-right panel of Fig.~\ref{conR} that the
affine scalar curvature $R$ remains constant, $6m/b$. The behavior
of the torsion $\Phi$ can be understood through Eq.~(\ref{dtphi}).
$\Phi$ will increase (or decrease) until its value balances the rhs
of Eq.~(\ref{dtphi}); this mainly depends on the sign change of the
term $3H\Phi$ provided $H>0$ and ${\cal T}>0$. With the current
initial choice in this case, $\Phi$ decreases promptly at present
until the balancing point is reached, as seen in the bottom-left
panel of Fig.~\ref{conR}. However, $\Phi$ will not be a constant
since the rhs of Eq.~(\ref{dtphi}) still changes with time. The
Hubble function $H$ will always decrease, and approach to the fixed
value $m/\sqrt{2{\bar a}_1b}\approx 0.84$, as shown in the
upper-left panel of Fig.~\ref{conR}, since the rhs of
Eq.~(\ref{dtHconR}) is always negative. The acceleration of the
expansion factor, $\ddot a$, is positive at late time, as seen in
the upper-right panel of Fig.~\ref{conR}.

It is very interesting to see how the universe evolves if the scalar
affine curvature $R$ has a tiny deviation from the constant value
$6m/b$. Therefore, we chose the initial values of $R$ as
$R(t=1)=6m/b-10^{-8}$ and $R(t=1)=6m/b+10^{-8}$ and evolved the
system while keeping all the other initial choices the same as in
the $R=6m/b$ case. The results are also plotted in Fig.~\ref{conR}.
In Fig.~\ref{conR}, the (blue) dashed lines are for the
$R(t=1)=6m/b-10^{-8}$ case and the (red) dot-dashed lines are for
the $R(t=1)=6m/b+10^{-8}$ case. The results show that once the
scalar affine curvature $R$ is smaller than $6m/b$ by a tiny amount,
the values of $R$ and thus the other fields will eventually return
to a damped oscillating mode. Therefore, the universe will
eventually approach a static condition. On the other hand, if the
scalar affine curvature $R$ is bigger than $6m/b$ by a tiny amount,
the values of $R$ and thus the other fields will rise unboundly.
Either way the affine curvature will never recover its constant
value. Therefore, the constant curvature case represents an {\it
unstable} universe with an effective cosmological constant or a
negative-kinetic-energy field. This phenomenon demonstrates the
inherent instability of a system with a negative kinetic energy.
However, as long as the deviation of $R$ from $6m/b$ is small
enough, it would be difficult from Fig.~\ref{conR} to predict the
future of the universe, since all the lines of these three cases are
virtually overlapped together until a very late time.  By careful
fine tuning we can arrange for a large variety of outcomes.  This
``chaotic'' behavior well illustrates just how we can lose all
physical predictability if we allow such {\em unphysical} parameter
choices.

Although the above parameter choice has the virtue of explaining the
accelerating expansion of the Universe and the cosmological
constant, we cannot accept such a parameter choice
here since it violates the fundamental assumption of the
positivity of the kinetic energy. Therefore, in the following cases,
we will return to our normal {\em physical} assumption, i.e.,
$a_1>0$ and $\mu>0$.

%%%%%%%%%%%%%%%%%%%%%%%%%%%%%%%%%%%%%%%%%%%%%%%%%%%%
\subsection{Case II: oscillating acceleration of $a$}
%%%%%%%%%%%%%%%%%%%%%%%%%%%%%%%%%%%%%%%%%%%%%%%%%%%%
For this case, we take the initial values of the field to be
\begin{eqnarray}
  &&a(0)=10,\quad H(0)=5\times 10^{-3},\nonumber\\
  &&\Phi(0)=2\times 10^{-4},\quad R(0)=-2\times 10^{-3},\nonumber
\end{eqnarray}
and the parameters are taken to be
\[
  \mu=1.2,\quad b=4,
\]
The results plotted in Fig.~\ref{oscR1} show that $\ddot a$, $\Phi$,
and $R$ are damped periodic.

In particular $R$ has a periodic character as shown in the
bottom-right panel of Fig.~\ref{oscR1}. According to
Eq.~(\ref{anaT}) its period is
$T=2\pi\sqrt{-a_1b/2a_0\mu}\approx3.63$, which is close to the
period of the variables shown in Fig.~\ref{oscR1}. The most
interesting part is the behavior of $\ddot a$, which is periodic
with the same period as $\Phi$ and $R$. As shown in the top-right
panel of Fig.~\ref{oscR1}, $\ddot a$ could be positive as well as
being negative and the pattern of its function is similar to the
pattern of $R$. Therefore the behavior of $H$ is a declining
baseline plus a damped oscillation, as shown in the top-left panel
of Fig.~\ref{oscR1}.

On a broader viewpoint of the evolution of this system, $\ddot a$,
$\Phi$, and $R$ will be slowly damped, and $H$ will approach zero
after a long time.  The important feature of this case is that the
universe could oscillate due to dynamic torsion. In such a scenario
the present day acceleration of the Universe is not so strange,
$\ddot a$ is oscillating and it happens to be increasing at this
time. Furthermore, the oscillation period would be determined mainly
by the parameters and the initial values of the fields in this
model. This encouraged us to try to find parameter values and
initial conditions which more nearly resemble the current status of
the Universe. Such a choice it will be shown in the next case.

%%%%%%%%%%%%%%%%%%%%%%%%%%%%%%%%%%%%%%%%%%%%%%%%%%%%%%%%%%%%%%%%%%%%%%
\subsection{Case III: A presently accelerating universe}
%%%%%%%%%%%%%%%%%%%%%%%%%%%%%%%%%%%%%%%%%%%%%%%%%%%%%%%%%%%%%%%%%%%%%%
\begin{figure*}[thbp]
\begin{tabular}{rl}
\includegraphics[width=8.8cm]{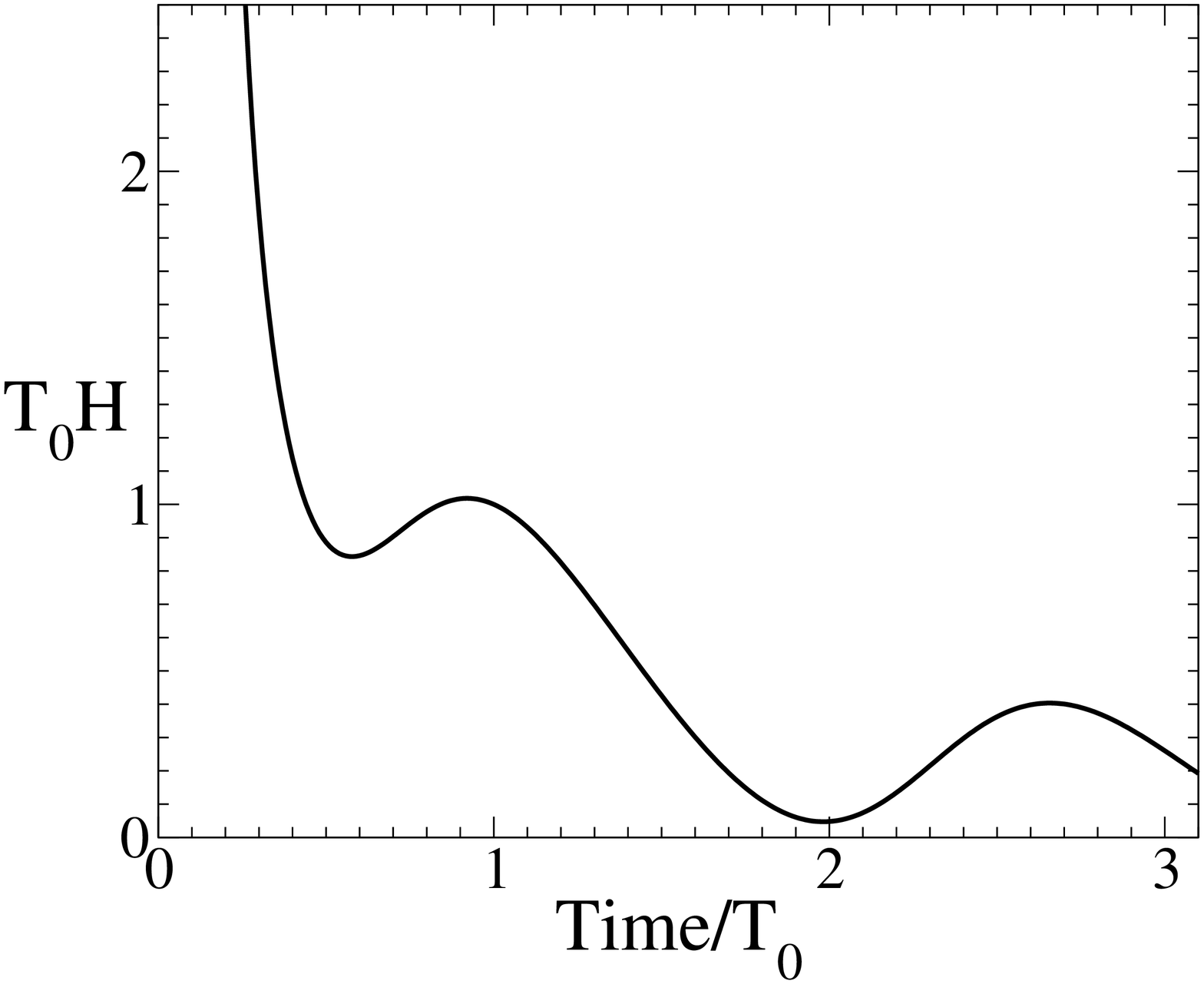}&
\includegraphics[width=8.8cm]{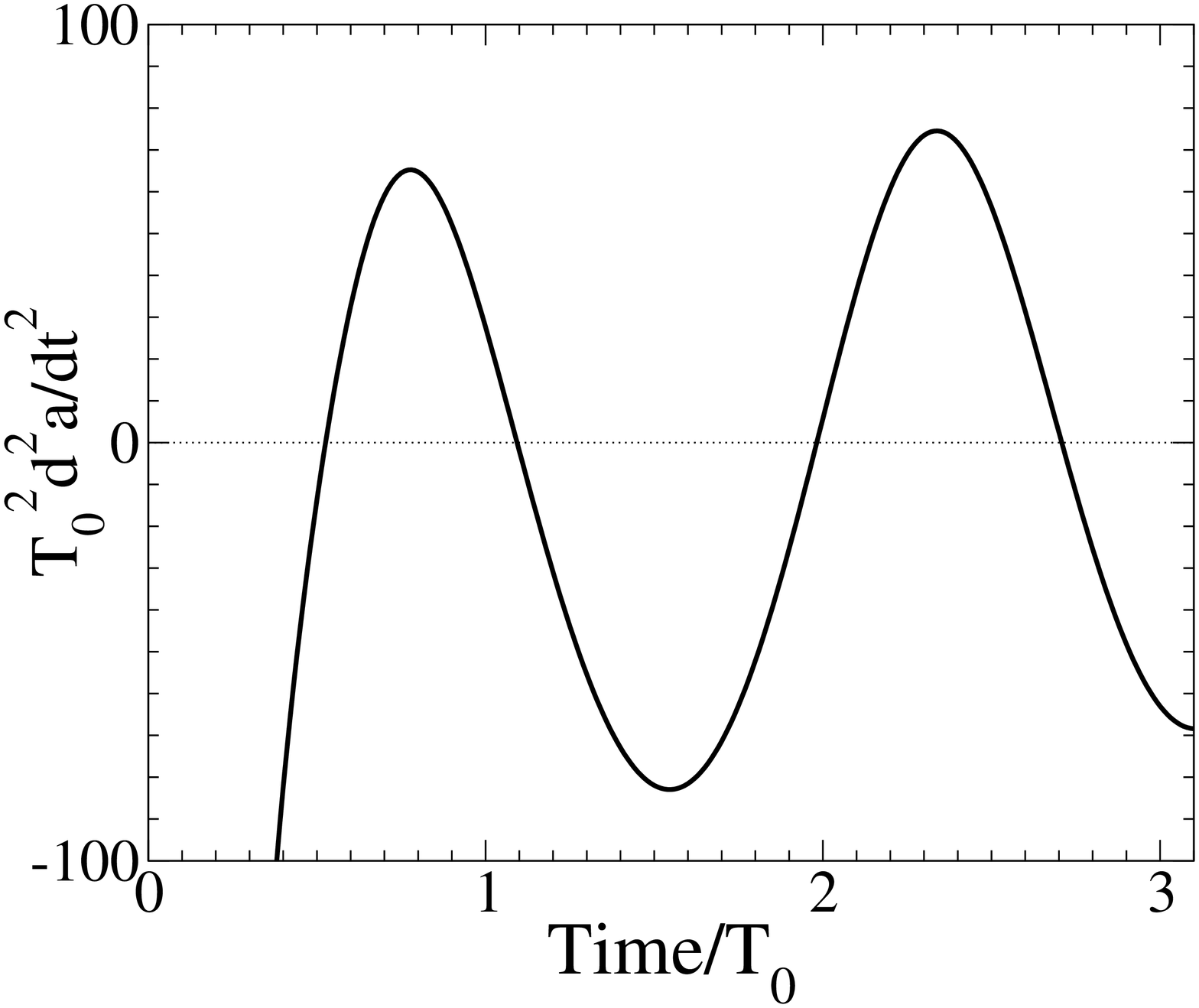} \\
\includegraphics[width=8.8cm]{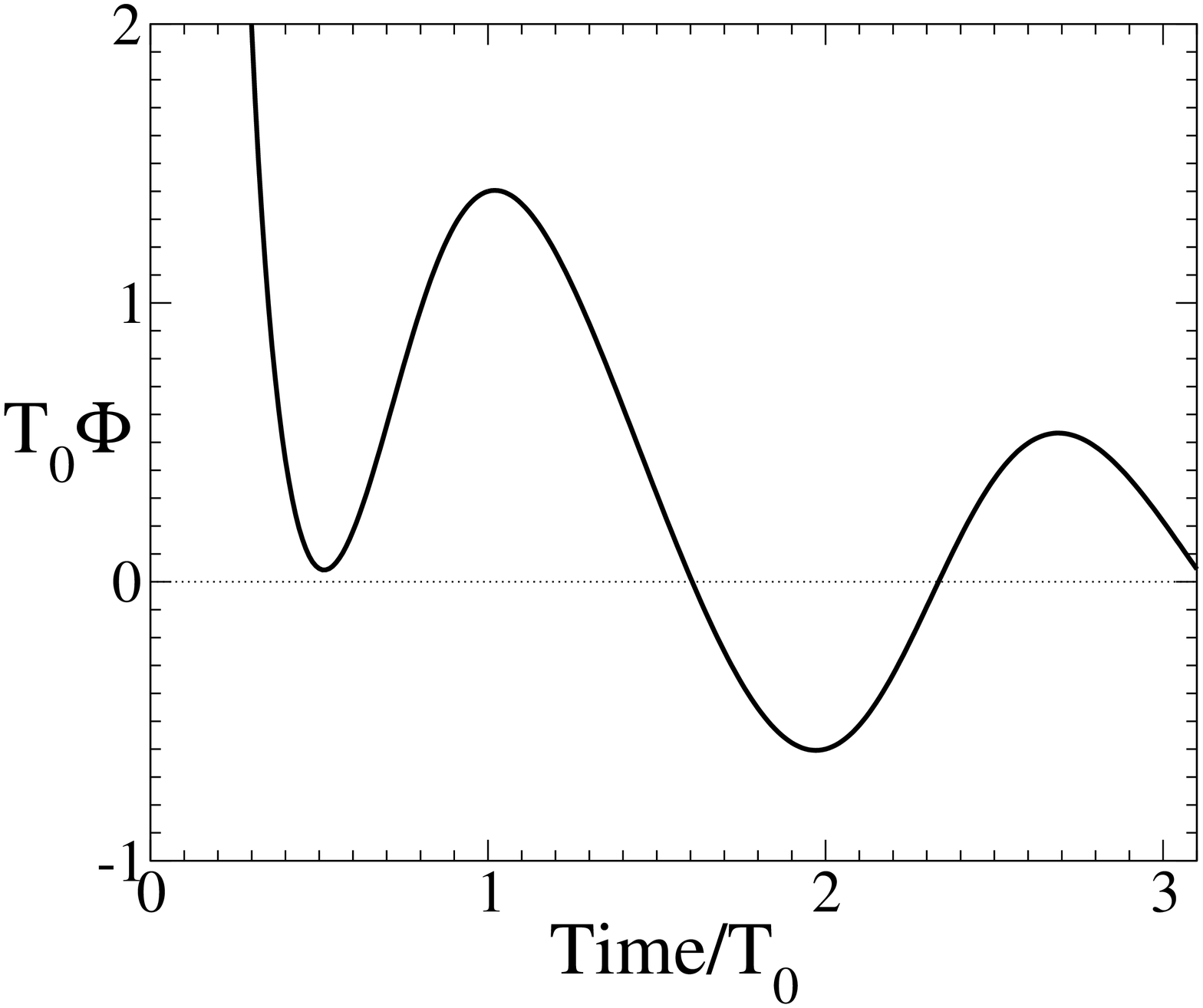}&
\includegraphics[width=8.8cm]{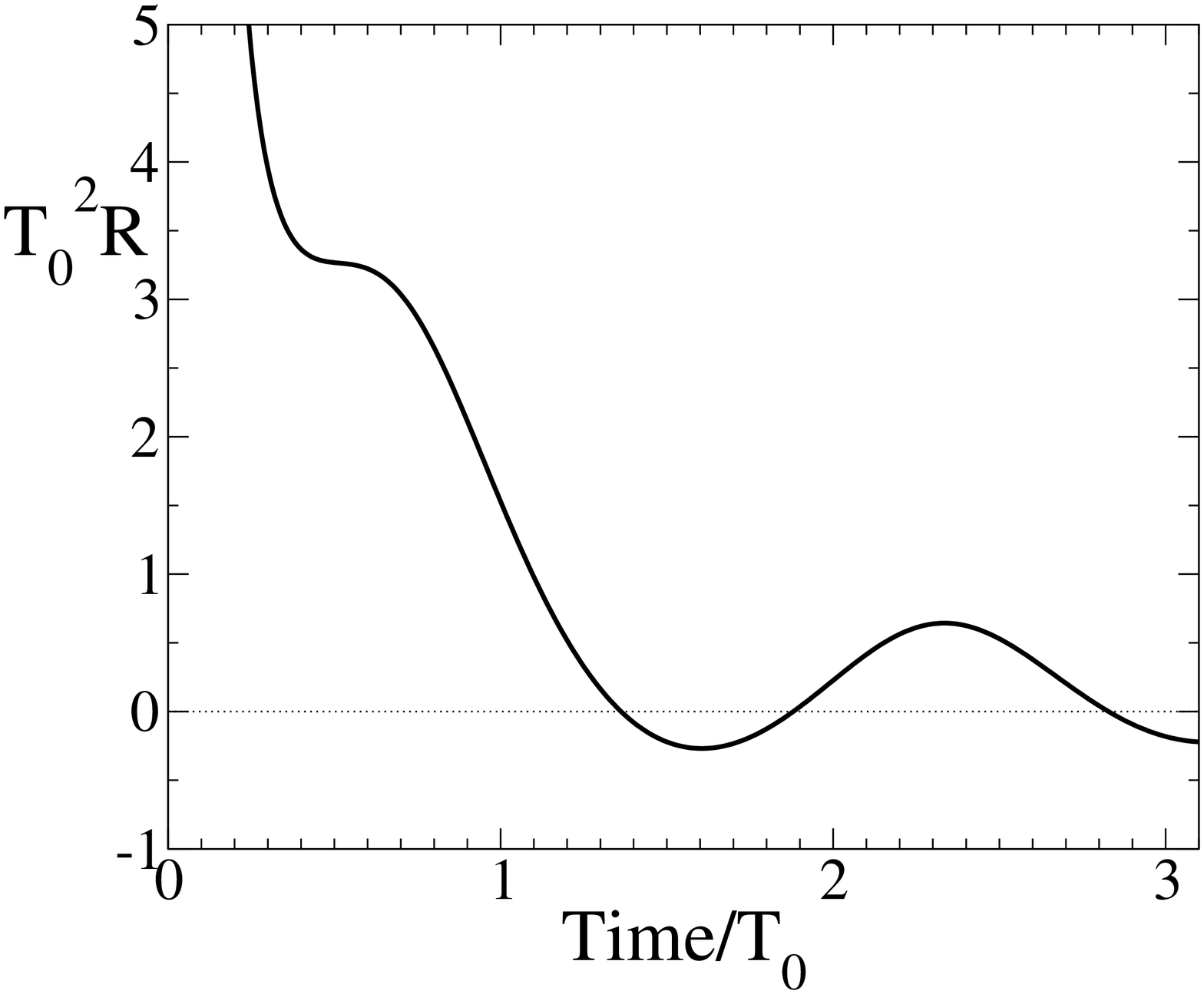}\\
\includegraphics[width=8.8cm]{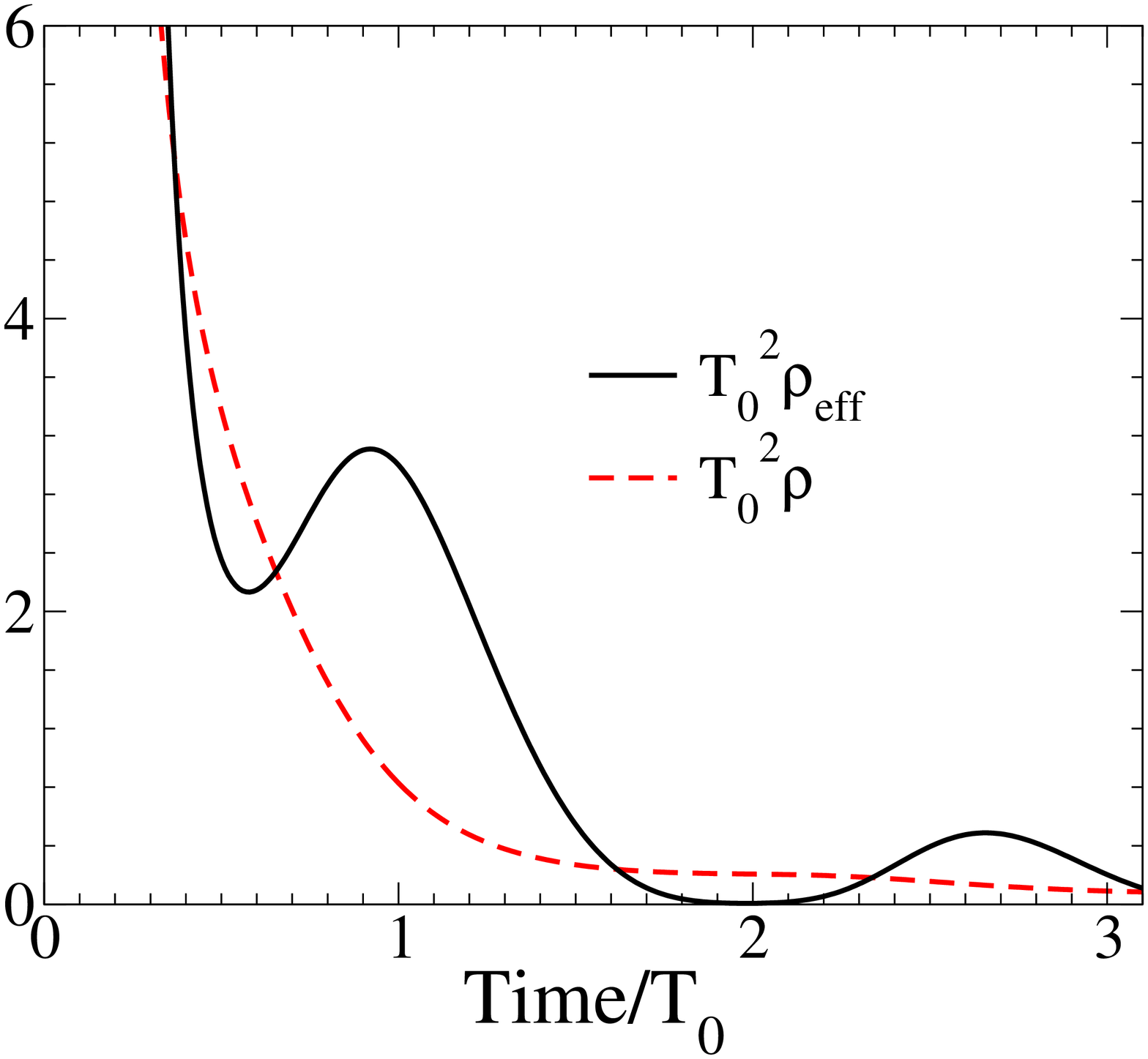}&
\includegraphics[width=8.8cm]{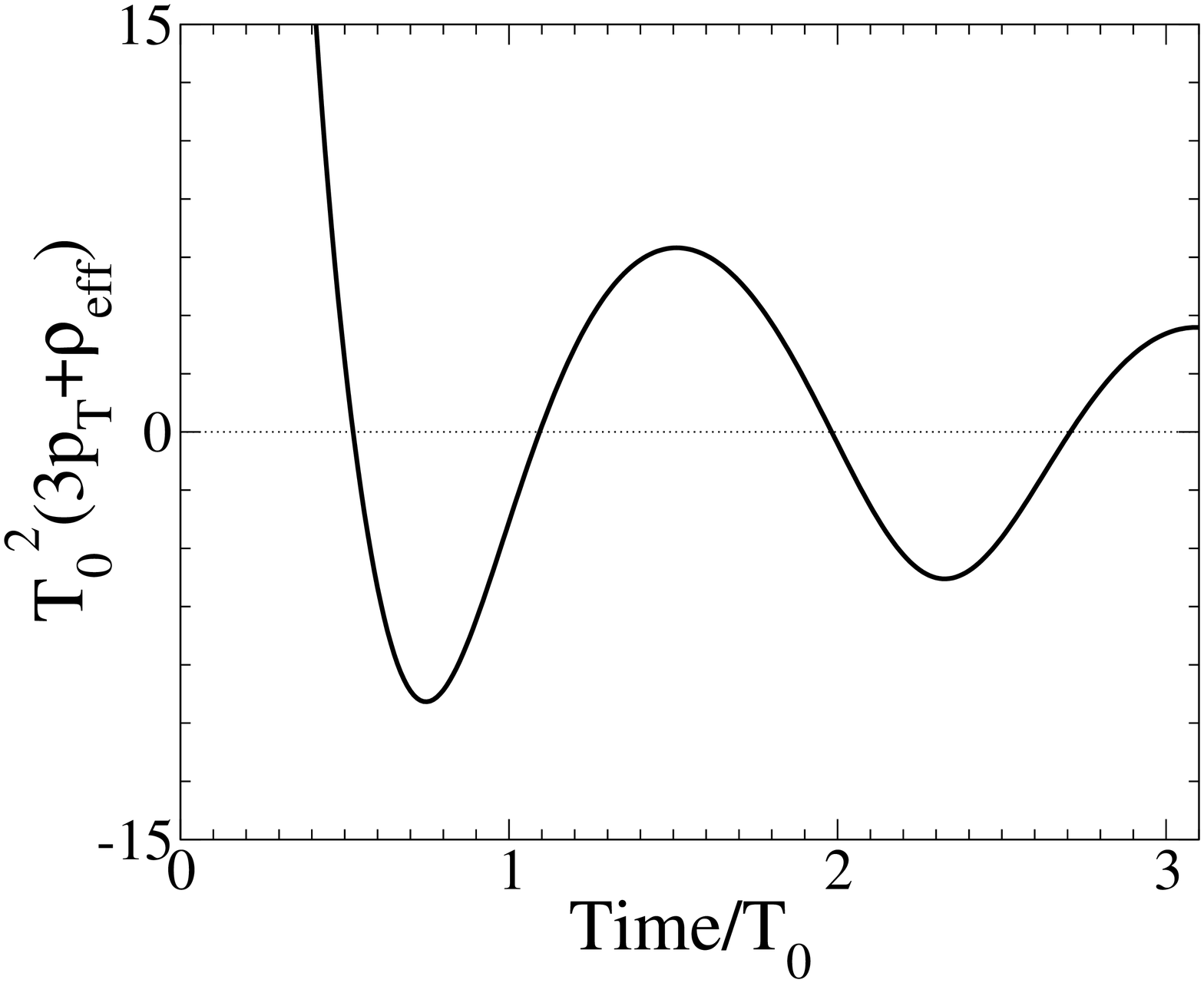}
\end{tabular}
\caption{Evolution of the Hubble function, $H$, the 2nd time
derivative of the expansion factor, $\ddot a$, the temporal
component of the torsion, $\Phi$, the affine scalar curvature, $R$,
the mass density, $\rho$, the effective mass density, $\rho_{\rm
eff}\equiv\rho+\rho_{\rm T}$, and the quantity $3p_{\rm T}+\rho_{\rm
eff}$, as functions of time with the parameter choice and the
initial data in Case III.} \label{case3}
\end{figure*}

In this case, we would like to compare the numerical values of the torsion model
with the observational data of the universe.
The initial data is set at the current time
$t_0=1$, after scaling, instead of $t_0=0$.
The parameters and initial conditions chosen are as follows
\begin{eqnarray}
  &&a(t_0=1)=50,\quad H(t_0=1)=1,\quad \Phi(t_0=1)=1.4,\nonumber\\
  &&Y(t_0=1)\equiv R(t_0=1)+\frac{6\mu}{b}=6.2,\nonumber
\end{eqnarray}
and
\[
  \mu=1.09,\quad b=1.4.
\]
Here the initial data has been scaled according to
Eqs.~(\ref{scale1}--\ref{scale2}) such that the current value of the
Hubble function is unity. Therefore we get realistic values in our
universe: the Hubble constant at present, $H(t_0=1)$, is
\begin{equation}
H=\frac{1}{4.41504\times 10^{17}}\cdot\frac{1}{\rm s}\approx 70
\frac{\rm km}{{\rm s}\cdot{\rm Mpc}}\,.
\end{equation}
The results of the evolution with the parameters and initial
conditions are plotted in Fig.~\ref{case3}. In the top-left panel
the Hubble function $H$ is damped-oscillating at late time. In the
top-right panel, it is obvious that $\ddot{a}$ is damped and
oscillating during the evolution and is positive at the current time
$t\approx 1$, which means the expansion of the universe is currently
accelerating. $\Phi(t)$ and $R(t)$ are also plotted in
Fig.~\ref{case3} to show the correlation of the evolution between
these variables. We observe that the values of the variables $H(t)$,
$\ddot a(t)$, $R(t)$, and $\Phi(t)$ become relatively high before
$t/T_0=0.4$. However this situation need not be taken too seriously,
since it describes the earlier time of the universe, and our
matter-dominated era assumption is not appropriate for such an early
period of time.

In order to have a deeper understanding of the settings of this
case, the matter density $\rho$, the effective mass density
$\rho_{\rm eff}=\rho +\rho_{\rm T}$, and the quantity $3p_{\rm T}
+\rho_{\rm eff}$ are plotted in the bottom panels of
Fig.~\ref{case3}. The value of $\rho$, shown in the bottom-left
panel, is decreasing at $t\approx 1$ as the universe is expanding
and is always positive, while the effective mass density $\rho_{\rm
eff}$, plotted in the same panel, shows an ``oscillating'' behavior
around the curve of $\rho$. The oscillating behavior of $\rho_{\rm
eff}$ comes from the contribution of the torsion-induced mass
density $\rho_{\rm T}$ and simply indicates that $\rho_{\rm T}$ is
not positive-definite in general. In fact the value of $\rho_{\rm
T}$ turns from negative to positive when the time is around
$t\approx0.7$. As to the quantity $3p_{\rm T} +\rho_{\rm eff}$, we
can understand its importance for the evolution of the universe
through Eq.~(\ref{ddota}) in which the value of this quantity
decides the status of the acceleration. We can see this much more
clearly by checking the correlation between the curves of $\ddot a$
and $3p_{\rm T} +\rho_{\rm eff}$ in Fig.~\ref{case3}. Also by
comparing the two bottom panels of Fig.~\ref{case3}, it is obvious
that the torsion-induced pressure $p_{\rm T}$ is negative when the
universe accelerates, and positive when the universe decelerates.

In this case the scaled value of $\rho(t=1)=0.83$ and its physical
value is $\rho(t=T_0)=2.61\times 10^{-30}{\rm g}/{\rm cm}^3$. The
Universe is supposed to be very close to the critical density,
$\rho_c\equiv3c^2H^2/8\pi G=9.47\times 10^{-30}{\rm g}/{\rm cm}^3$;
we find the ratio $\Omega_{\rm m}\equiv\rho/\rho_c=28\%$. In the
standard $\Lambda$CDM model, $\Omega_{\rm m}\sim30\%$ with $5\%$
baryonic matter and $25\%$ dark matter. For our model $\Omega_{\rm
T}\equiv\rho_{\rm T}/\rho_c=72\%$ acts like the energy density of dark
energy. Therefore, this torsion model is able to describe a
presently accelerating expansion of the Universe with a proper
amount of matter density. From the field equations  we can see that
the {\it effect} of the ``dark energy" mainly comes from the
nonlinearity of the field equation driven by the dynamic scalar
torsion.

\begin{figure*}[thbp]
\begin{tabular}{c}
\includegraphics[width=17.8cm]{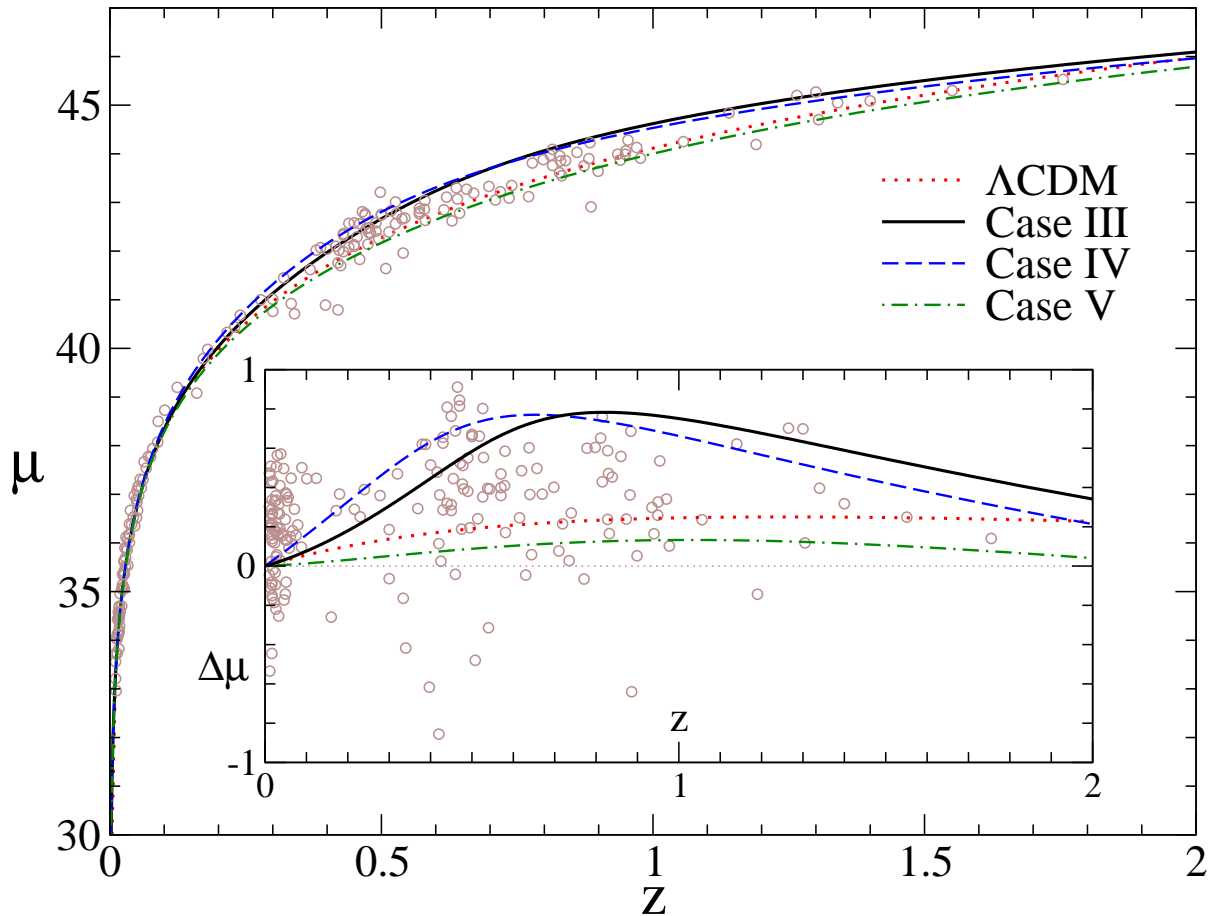}
\end{tabular}
\caption{Comparison of different torsion models and the standard
$\Lambda$CDM model with the observational data via the relation
between the distance modulus $\mu$ and the redshift $z$. The
supernovae data points, plotted with (brown) circles, come from
\cite{Reis04}. the result of standard $\Lambda$CDM model
($\Omega_{\rm m}=0.3$, $\Omega_\Lambda=0.7$) is plotted by the bold
(red) dotted line. The results of Case III, IV, V, are represented
by the bold solid line, the (blue) dashed line, and the (green)
dot-dashed line, respectively. In the inset, the models and data are
shown relative to an empty universe model ($\Omega=0$).}
\label{case345}
\end{figure*}

\begin{table}[thbp]
\begin{tabular}{cccccccccc}
\hline
Case &$\mu$&$b$&$H(1)$&$\Phi(1)$&$Y(1)$&$a(1)$&$\ddot{a}(1)$
&$\displaystyle{\frac{\rho(1)}{10^{-30}{\rm g}/{\rm cm}^3}}$\\
\hline
III& 1.09 & 1.4 & 1 & 1.4 &  6.2 & 50 & 27.59 & 2.61 \\
IV & 1.27 & 1.1 & 1 & 0.8 & 11.3 & 50 & 70.29 & 5.23 \\
V  & 1.38 & 1.1 & 1 & 1.1 &  9.9 & 50 &  4.57 & 2.48 \\
\hline
\end{tabular}
\caption{Here the parameter $a_0$ is set to be $1$ in all of the
three cases; $H(1)$ means $H(t={\rm now})$, $\Phi(1)$ means
$\Phi(t={\rm now})$, etc, under the scaling
Eqs.~(\ref{scale1}--\ref{scale2}).} \label{allpar}
\end{table}

%%%%%%%%%%%%%%%%%%%%%%%%%%%%%%%%%%%%%%%%%%%%%%%%%%%%%%%%%%%%%%%%%%%%%%
\subsection{Other Cases}
%%%%%%%%%%%%%%%%%%%%%%%%%%%%%%%%%%%%%%%%%%%%%%%%%%%%%%%%%%%%%%%%%%%%%%
\label{case45}
We continue to look at two more cases, which are listed in
Table~\ref{allpar} along with Case III, obtained by taking different
values of the parameters and the initial conditions, along with
physical values of the significant mass density $\rho$. We find that
the results of the other two cases have a behavior qualitatively
similar to that of Case III.

Now we would like to compare our results with the supernovae data.
Distance estimates from SN Ia light curves are derived from the luminosity
distance
\begin{equation}
d_L\equiv\sqrt{\frac{L_{\rm int}}{4\pi{\cal F}}}=cT_0 a(1)(1+z)
\int^t_1\frac{{\rm d}t}{a(t)}\,,
\end{equation}
where $L_{\rm int}$ and $\cal F$ are the intrinsic luminosity and
observed flux of the SN, and the redshift $z\equiv a(1)/a(t)-1$.
Logarithmic measures of the flux (apparent magnitude, $m$) and
luminosity (absolute magnitude, $M$) were used to derive the
predicted distance modulus \cite{munotmu}
\begin{equation}
\mu=m-M=5\log_{10}d_L+25\,,
\end{equation}
where $m$ is the flux (apparent magnitude), $M$ is the luminosity
(absolute magnitude), and $d_L$ in the formula should be in units of
megaparsecs. We found the relations between the predicted distance
modulus $\mu$ and the redshift $z$ in the three cases; they are
plotted in Fig.~\ref{case345}. For comparison, we also plot the
prediction of the $\Lambda$CDM model with $\Omega_{\rm m}=0.3$ and
$\Omega_{\Lambda}=0.7$ by employing the following formula
\cite{Reis04}
\begin{equation}
d_L=cT_0(1+z)\int_0^z\frac{{\rm d}z}{\sqrt{(1+z)^2(1+
\Omega_{\rm m}z) -z(2+z)\Omega_\Lambda}}\,.
\end{equation}
The astronomical observational data \cite{Reis04,SNIaRP} are also
plotted in Fig.~\ref{case345} for comparison. The plots show that
for small redshift $z$ (e.g., $z<1.9$) all three cases of the
dynamical torsion models give an accelerating universe just like the
$\Lambda$CDM model does. For larger $z$ these cases might turn the
Universe into a deceleration mode, which is consistent with the
behavior of the various quantities shown in Fig.~\ref{case3}. We can
see that Case V gives the closest curve behavior to the one from the
$\Lambda$CDM model, although in Case V the matter density is only
about $26\%$ of the critical density. However, it was not meant to
have a detailed comparison in this plot between our models with the
$\Lambda$CDM model. Instead, in Fig.~\ref{case345}, we demonstrate
the possibility of the scalar torsion field accounting for the
effect of dark energy with a suitable set of parameters and initial
data. This allows us to study the dark energy problem from a new and
different angle.
%%%%%%%%%%%%%%%%%%%%%%%%%%%%%%%%%%%%%%%
\section{Discussion}
%%%%%%%%%%%%%%%%%%%%%%%%%%%%%%%%%%%%%%%%
In this work we introduce into the evolution of a universe without a
cosmological constant a certain dynamical PGT scalar torsion mode
taken from our earlier work \cite{YN99}. From the assumption of the
homogeneity and isotropy of the universe, only the temporal
component of the torsion $\Phi$ will survive and affect the
evolution of the universe at late times. With the field equations
(\ref{dta}--\ref{dtR}), we analyzed analytically and numerically the
evolution of the system. We found that in generic cases, i.e.,
$R+6\mu/b\ne0$, with the proper parameter choice (i.e., $a_1>0$ and
$\mu>0$), the torsion $\Phi$ in the system is dynamic, and $\ddot
a$, $\Phi$, $R$ tend to have a damped periodic behavior with the
same period while the behavior of $H$ is a declining baseline plus a
damped oscillation. With certain choices of the parameters of $\mu$
and $b$, and of the initial data of $H$, $\Phi$, and $R$, like Cases
III--V in the previous section, this model can describe an
oscillating universe with an accelerating expansion at the present
time.

Before we can give an adequate discussion of the viability of this
model as an explanation of the accelerating universe, we should
check whether this model can survive under the constraints of the
theoretical and experimental tests.

There have been numerous investigations on the existence of torsion
since this geometric quantity entered the realm of gravity (see
\cite{HaRT02,ShIL02,AHPJ04} and the references therein). As
mentioned above, this model has not only passed the important
classical tests (``no-ghosts" and ``no-tachyons"), it is also one of
the two scalar torsion modes---the only PGT cases which are known to
have a well posed initial value problem \cite{YN99} and which may
well be the only viable dynamic PGT torsion modes that can evade the
non-linear constraint problems.   There have also been some
laboratory tests in search of torsion \cite{CTNW93,NiWT96}. The main
idea among these experiments is the spin interaction between matter
and torsion. The cosmological tests on torsion investigate the
effect of torsion-induced spin flips of neutrinos in the early
Universe which could alter the helium abundance and have other
effects on the early nucleosynthesis \cite{CILS99,BruM99}. However
Dirac fermions interact only with the totally antisymmetric
pseudo-scalar part of the torsion. Thus these tests can only
consider the pseudo-scalar mode (axial-vector torsion), not the
scalar mode torsion used in our model. The type of torsion used in
our model does not interact directly with any known matter. Thus,
these tests cannot really give a serious constraint on the amplitude
of our scalar-mode torsion.

Among the models in which torsion is applied to the cosmological
problem, Capozziello {\it et al.}~\cite{CaCT03,CSet03} have done a
serious study on replacing the role of the cosmological constant in
the accelerating Universe. With a totally antisymmetric torsion
without dynamical evolution, their model is consistent with the
observational data by tuning the amount of the torsion density,
although this model cannot solve the coincidence problem. On the
other hand, the oscillating universe models with a designed
mechanism: an oscillating potential, an oscillating parameter of the
equation of state, etc.~\cite{DoKS00,RSPC03,FLPZ06} aim to solve the
coincidence problem. Here we found that our model takes some virtues
from both kind of models, i.e., our model is capable of solving the
coincidence problem of an accelerating universe with a dynamical
scalar-mode torsion, which is {\it naturally} obtained from the
geometry of the Riemann-Cartan spacetime, instead of from an exotic
scalar field or a designed mechanism.

If we consider the spacetime as Riemannian instead of
Riemann-Cartan, by absorbing the contribution of the torsion of this
model into the stress-energy tensor on the rhs of the Einstein
equation, then this contribution will act as a source of the
Riemannian metric, effectively like an {\em exotic} fluid with its
mass density $\rho_{\rm T}$ and pressure $p_{\rm T}$ varying with
time (even though the time evolution of the torsion is not like that
of a such a fluid). Moreover, the effective fluid appears to have
presently a negative pressure, and consequently a negative parameter
in the effective equation of state, i.e., $\omega_{\rm T}\equiv
p_{\rm T}/\rho_{\rm T}$, which drives the universe into accelerating
expansion. Note that there is no constraint on the value of
$\omega_{\rm T}$ which appears here, and its value could vary from
time to time. It should be stressed that this is not a real physical
fluid situation; the truth is that $\omega_{\rm T}$ is nothing like
``a torsion field equation of state'', it is just a proportionality
factor between $\rho_{\rm T}$ and $p_{\rm T}$, two expressions which
effectively summarize the contribution of torsion acting as a source
of the metric. The ratio $\omega_{\rm T}$ is of interest only to
help understand the acceleration of this model and to enable a
limited comparison with other dark energy proposals.

One might be concerned about the value of the parameter $b$. Its
value should be small enough to be consistent with the
constraints on the affect of the quadratic order term $R^2$ on the
large scale structure of universe. The values of $b$ we choose,
i.e., $b/(a_0 T^2_0)$ in the conventional unit, are on the order of
unity. These chosen values are bigger than the magnitude of a
related parameter, estimated in \cite{kkos92}; however one cannot
expect that estimate to be applicable here---since in that work
quadratic {\em Riemannian curvature} terms were considered (they
lead to 4th order field equations) instead of the
affine curvature terms we have used (which give 2nd order
equations). As far as we know the parameter $\mu$ does not have too
much constraint on it, except for its positivity as a mass
parameter, since the baryonic matter will only interact with the
scalar torsion indirectly by gravitation.

One may wonder: how large must the torsion be in order to produce
observable effects in the the present day universe, e.g., the
observed acceleration? conversely, how large can the torsion be
without violating some observational constraint?  The questions
merit a detailed study.  Here is a simple argument that indicates a
magnitude. Let us compare the terms in the Lagrangian density and
the field equations for the PGT scalar torsion model and the
Einstein theory with a cosmological constant.   Note that the
presumed cosmological constant is ``so small'' that it has no
noticeable effect in the laboratory, nor on the solar system scale,
nor on the galactic scale.  Nevertheless it is large enough to have
the dominant effect on the cosmological scale. Hence we are led to
infer that we should consider $a_1T^2\sim bR^2 \sim \Lambda \sim a_0
\rho \sim H^2$.  With such a choice we can expect that torsion may
be able to accelerate the universe and yet not be conspicuous on
smaller scales.

The $0^+$ torsion mode in this model effectively gives a scalar
field, yet this scalar field is, in fact, quite different from the
various scalar field models of ``exotic matter'', e.g., the {\it
quintessence} models, in several significant ways: (i) torsion
cosmology is derived naturally from a geometric gravitational
theory, which is based on fundamental gauge principles, instead of
on the hypothesis of the existence of a dark  energy tailored to
producing an explanation of an accelerating universe; (ii) thus
there are only a couple of free parameters in torsion cosmology,
instead of an {\it ad hoc} potential that can be rather arbitrarily
chosen to fit the observations. Therefore, a torsion cosmological
model should be more restrictive, and should be easier to be
confirmed or falsified; (iii) based on its tensorial character, the
coupling of torsion to the other fields is nothing like that which
has ever been advocated for hypothetical scalar fields. Consequently
we see no way to simply replace the scalar mode torsion with an
effectively equivalent quintessence model. Thus torsion cosmology
and the quintessence models are characteristically different, even
though there are some similarities.

Due to its intriguing behavior, we also turned our attention to a
degenerate case, $R=-6\mu/b$, in Case I with the relaxed parameter
choice of $a_1<0$ and $\mu<0$ instead of the normal choice $a_1>0$
and $\mu>0$. Although such a choice is against the positivity of
kinetic energy, we explored this scenario since it could mimic the
cosmological constant and the other cosmological models with a
negative kinetic energy. Indeed the result does show an accelerating
universe at late time. However, our further numerical experiments
also show that in such a case it describes a very unstable universe.
A small perturbation from the constant curvature will cause a sudden
change which would only become apparent at some time in the future.
%%%%%%%%%%%%%%%%%%%%%%%%%%%%%%%%%%%%%%%
\section{Conclusion}
%%%%%%%%%%%%%%%%%%%%%%%%%%%%%%%%%%%%%%%%
In this work we considered the scalar torsion mode of the PGT on in
a cosmological setting and proposed it as a viable model for
explaining the current status of the Universe. Besides having a
better understanding of the PGT, we study the prospects of
accounting for the outstanding present day mystery---the
accelerating universe---in terms of an alternate gravity theory,
more particularly in terms of the PGT dynamic torsion. With the
usual assumptions of isotropy and homogeneity in cosmology, we find
that, under the model, the Universe will oscillate with generic
choices of the parameters. The torsion field in the model could play
the role of dark energy. With a certain range of parameter choices,
it can account for the current status of the Universe, i.e., an
accelerating expanding universe with a value of the Hubble constant
which is approximately the present one. Thus we have considered the
possibility that a certain geometric field, dynamic scalar
torsion---which is naturally expected from spacetime gauge
theory---could fully account for the accelerated universe.

The source of the torsion could come indirectly from the huge
density of the particles with sufficient spin  alignment in the
early universe. This scalar mode of torsion could be considered as a
``phantom" field, at least in the matter-dominated epoch, since it
will not interact directly with matter; it only interacts indirectly
via gravitation. Then the dynamics of the scalar torsion mode could
drive the Universe in an oscillating fashion with an accelerating
expansion at present.  It is quite remarkable that a gauge theory of
dynamic geometry naturally presents us with such a ``phantom''
field.  This natural geometric field could act like a dark energy.

However, there are also some points which need to be studied in much
more detail before this model can more closely conform to reality.
The model in Cases III--V of the previous section, suggests that the
mass parameter of the torsion, $\mu$, might be close to $a_0$, and
the parameter for the ``kinetic'' energy density of the torsion,
$b$, may need to be as huge as $T_0^2$ to achieve an accelerating
universe. The restricted window of the parameter choices which
allows a behavior like that of our universe might render the model
less favored, even though the matter in the universe is not able to
directly interact with the torsion. Meanwhile, the required choice
of initial data and the values of the parameters may make this model
unsuited to solving the fine-tuning problem.

These dark sides should not be able to diminish the possibility of
the scalar mode of the torsion in this model playing a significant
role in the the evolution of the Universe. The model has only a few
adjustable parameters, so scalar torsion may be easily
falsified---as ``dark energy.''  If it turns out that the
accelerated universe cannot be explained in this way---that
something else has the dominant dark energy role---it would still be
reasonable to expect that there may be some observable cosmological
effects from dynamic scalar torsion. Also, here we only used one of
the viable modes of torsion in PGT; the our model will be more
general if it is extended to include all the viable PGT torsion
modes. We believe that future investigations along this line should
be open to these possibilities.

%%%%%%%%%%%%%%%%%%%%%%%%%%%%%%%%%%%%%%%%
\section*{Acknowledgments}
%%%%%%%%%%%%%%%%%%%%%%%%%%%%%%%%%%%%%%%%
The authors are grateful to Chopin Soo for his helpful suggestions
and discussions. This work was supported in part by the National
Science Council of the R.O.C. (Taiwan) under grant
Nos.~NSC94-2112-M-006-014, NSC95-2112-M-006-017-MY2 and NSC
95-2119-M008-027. This work was also supported in part by the
National Center of Theoretical Sciences and the (NCU) Center for
Mathematics and Theoretical Physics. Some of the calculations were
performed at the National Center for High-performance Computing in
Taiwan.
%%%%%%%%%%%%%%%%%%%%%%%%%%%%%%%%%%%%%%%%


\begin{thebibliography}{99}
\bibitem{0minus} This ``scalar potential'' can be identified as a
non-linear function of the pseudoscalar curvature,
as can be seen from Eq. (3.14) in \cite{HNZ96}. If one tried to
introduce a scalar potential for the axial torsion directly into the
PGT action it would lead to a different set of dynamic equations.
Because the quadratic curvature terms effectively include derivatives
of the torsion, it thus would have second derivatives of the scalar
potential---which would give rise to an undesired fourth order
equation for the potential.
\bibitem{highord} Making this translation in all of the terms of the equations would
lead to relations with undesirable 3rd derivatives of the metric
(from the derivative of the curvature).  In what we are regarding as
an Einstein equation with an effective source there would be 2nd
derivatives of the metric not only on the lhs but also unpleasently
appearing (quadratically!) on the rhs in what is supposed to be the
effective source.
\bibitem{munotmu}
The symbol $\mu$ for the distance modulus has been used widely in the
literature and is well-understood. However, it has nothing to do with
the mass parameter $\mu$ defined in Sec. \ref{secpgt}. The meaning of $\mu$
is clear in the context since $\mu$ only represents the distance modulus
in the subsection, i.e., Sec. \ref{case45}.
\end{thebibliography}
\end{document}